\documentclass[]{pasj02} 
\usepackage[switch,mathlines]{lineno}
\usepackage{natbib} 

\jyear{2024}
\Received{2025/02/03}
\Accepted{2025/03/11}

\begin{document} 

\title{Outflowing photoionized plasma in Circinus X-1 using the high-resolution
X-ray spectrometer \textit{Resolve} onboard XRISM and the radiative transfer code
\texttt{cloudy}}

\author{
 Masahiro Tsujimoto,\altaffilmark{1}\altemailmark\orcid{0000-0002-9184-5556} 
 \email{tsujimoto.masahiro@jaxa.jp} 
 Teruaki Enoto,\altaffilmark{2}\orcid{0000-0003-1244-3100}
 Mar\'ia D\'iaz Trigo,\altaffilmark{3}\orcid{0000-0001-7796-4279}
 Natalie Hell,\altaffilmark{4}\orcid{0000-0003-3057-1536}
 Priyanka Chakraborty,\altaffilmark{5}\orcid{0000-0002-4469-2518}
 Maurice A. Leutenegger,\altaffilmark{6}\orcid{0000-0002-3331-7595}
 Michael Loewenstein,\altaffilmark{6}\orcid{0000-0002-1661-4029}
 Pragati Pradhan,\altaffilmark{7}\orcid{0000-0002-1131-3059}
 Megumi Shidatsu,\altaffilmark{8}\orcid{0000-0001-8195-6546}
 Hiromitsu Takahashi,\altaffilmark{9}\orcid{0000-0001-6314-5897}
 Tahir Yaqoob\altaffilmark{6,10,11}
 }
 \altaffiltext{1}{Japan Aerospace Exploration Agency (JAXA), Institute of Space and
 Astronautical Science (ISAS), Chuo-ku, Sagamihara, Kanagawa 252-5210, Japan}
 \altaffiltext{2}{Department of Physics, Graduate School of Science, Kyoto University,
 Sakyo-ku, Kyoto, Kyoto 606-8502, Japan}
 \altaffiltext{3}{European Southern Observatory, Garching bei M\"{u}nchen, Germany}
 \altaffiltext{4}{Lawrence Livermore National Laboratory, Livermore, CA 94550, USA}
 \altaffiltext{5}{Center for Astrophysics, Harvard \& Smithsonian, Cambridge, MA 02138, USA}
 \altaffiltext{6}{NASA / Goddard Space Flight Center, Greenbelt, MD 20771, USA}
 \altaffiltext{7}{Department of Physics, Embry-Riddle Aeronautical University, Prescott, AZ 86301, USA}
 \altaffiltext{8}{Department of Physics, Ehime University, Matsuyama, Ehime 790-8577, Japan}
 \altaffiltext{9}{Department of Physics, Hiroshima University, Higashi-Hiroshima, Hiroshima 739-8526, Japan}
 \altaffiltext{10}{Center for Space Sciences and Technology, University of Maryland, Baltimore County
 (UMBC), Baltimore, MD 21250 USA}
 \altaffiltext{11}{Center for Research and Exploration in Space Science and Technology,
 NASA / GSFC (CRESST II), Greenbelt, MD 20771, USA}
\altaffiltext{12}{Department of Astronomy, University of Maryland, College Park, MD 20742, USA}
 \KeyWords{radiative transfer --- stars: neutron --- techniques: spectroscopic --- X-rays: binaries}  
\maketitle

\begin{abstract}
 High-resolution X-ray spectroscopy is a key to understanding the mass inflow and
 outflow of compact objects. Spectral lines carry information about the ionization,
 density, and velocity structures through their intensity ratios and profiles. They are
 formed in non-local thermodynamic equilibrium conditions under the intense radiation
 field from the compact objects, thus radiative transfer (RT) calculation is a requisite
 for proper interpretations. We present such a study for a low-mass X-ray binary,
 Circinus X-1, from which the P Cygni profile was discovered using the
 X-ray grating spectrometer onboard Chandra. We observed the source using the X-ray
 microcalorimeter onboard XRISM at an orbital phase of 0.93--0.97 and revealed many
 spectral features unidentified before; the higher series transitions ($n \rightarrow
 1$; $n>2$) of highly-ionized (H- and He-like) S, Ca, Ar, and Fe in emission and
 absorption, the Fe K$\alpha$ and K$\beta$ inner-shell excitation absorption of
 mildly-ionized (O- to Li-like) Fe, and resolved fine-structure level transitions in the
 Fe Ly$\alpha$ and He$\alpha$ complexes. They blend with each other at different
 velocity shifts on top of apparently variable continuum emission that changed its flux
 by an order of magnitude within a 70~ks telescope time. Despite such complexity in the
 observed spectra, most of them can be explained by a simple model consisting of the
 photoionized plasma outflowing at $\sim$300~km~s$^{-1}$ and the variable blocking
 material in the line of sight of the incident continuum emission from the accretion
 disk. We demonstrate this with the aid of the RT code \texttt{cloudy} for the line
 ratio diagnostics and spectral fitting. We further constrain the physical parameters of
 the outflow and argue that the outflow is launched close to the outer edge of the
 accretion disk and can be driven radiatively by being assisted by the line force
 calculated using the RT simulation.
\end{abstract}


\section{Introduction}\label{s1}
Compact objects, such as black holes (BH) and neutron stars (NS), are a cosmic pumping
mechanism of mass, momentum, and energy. A part of the accreted matter onto the compact
objects, such as those in X-ray binaries and at the center of galaxies, is ejected in
the form of jets and winds. They carry momentum and energy concentrated in a fraction of
accreted mass, through which the redistribution of momentum and energy takes place in
the universe.

High-resolution X-ray spectroscopy is one of the best observational tools to unveil such an
outflowing mass. Because of the strong radiation field of compact objects, the matter
around them is often photoionized. Copious emission lines from the photoionized plasma,
as well as absorption lines imprinted in the continuum emission transmitted through it,
are routinely observed in such systems. They carry information on the ionization,
density, and velocity structures of matter around compact objects.

With the advent of the X-ray microcalorimeter (\textit{Resolve}; \citealt{ishisaki2022})
onboard the X-Ray Imaging and Spectroscopy Mission (XRISM; \citealt{Tashiro2020}), new
spectral features became available for diagnosing these photoionized plasmas. They are
particularly rich in the Fe K band (6--9~keV), which includes the Fe K$\alpha$ line from
neutral to low-ionized (up to Ne-like) ions, the inner-shell excitation lines from
mildly-ionized (F-like to Li-like) ions, fine-structure lines of He$\alpha$ and
Ly$\alpha$ ($n=2 \rightarrow 1$) line complexes from highly-ionized (He- and H-like)
ions, and their higher series counterparts ($n \rightarrow 1$; $n>2$).

The intensity ratios of these lines provide robust constraints for plasma
parameters. For this purpose, it is mandatory to consult the radiative transfer (RT)
calculations. In photoionized plasmas, the lines are formed in non-local thermodynamic
equilibrium (NLTE) conditions, in which radiative processes dominate the
heating/cooling, ionization/recombination, and excitation/deexcitation balances over
collisional processes. Matter and photons are only weakly coupled, and the radiation
field is far from blackbody. For the line ratio diagnostics, we need to know the charge
and level populations of matter by solving all the matter-photon interactions
consistently with the radiation field through numerical RT calculations.

In systems of our interest, two RT effects (continuum pumping and line optical depth)
are of particular importance. Depending on their combinations, they are labeled as case
A (lines are optically thin, no continuum pumping), B (optically thick, no continuum
pumping), C (optically thin, with continuum pumping), and D (optically thick, with
continuum pumping). See \citet{chakraborty2021} and references therein for more details.

The first is the continuum pumping, in which incident continuum photons radiatively excite
matter. Lines with a large oscillator strength are strongly enhanced in observed spectra
if a part of the incident emission is blocked in the line of sight, which is indeed
observed in Seyfert 2 galaxies \citep{kinkhabwala2002}, accretion disk corona sources
\citep{tsujimoto2024}, and eclipsing binaries during eclipse
\citep{wojdowski2003a}.

The second is the line optical depth. At the line center energy, the optical depth is given by
\begin{equation}
 \label{e01}
 \tau_{0} = \frac{1}{4\pi\epsilon_0} \frac{\pi e^{2}}{m_{\mathrm{e}}c} f
 \frac{1}{\sqrt{\pi}\Delta\nu_{\mathrm{D}}}
 N_{\mathrm{H}} A_{Z} A_{C} A_{L},
\end{equation}
in which $\Delta\nu_{\mathrm{D}} \equiv
\frac{E_0}{hc}\sqrt{\frac{2k_{\mathrm{B}}T}{m_{Z}} + v_{\mathrm{turb}}^{2}}$ is the line
width, $E_0$ is the energy of the line, $f$ is the oscillator strength, $A_{Z}$ is the
abundance of the element $Z$ relative to H, $A_{C}$ is the fractional charge population
of $Z^{C+}$, $A_{L}$ is the fractional level population of the level $L$, $m_{Z}$ is the
atomic mass of $Z$, $T$ is the plasma temperature, $v_{\mathrm{turb}}$ is the turbulent
velocity, and other symbols follow their conventions. To take the Fe \emissiontype{XXVI}
Ly$\alpha_1$ line ($^{2}P_{3/2} \rightarrow ^{2}S_{1/2}$) as an example, $E_0=6.973$~keV
\citep{yerokhin2019}, $f=0.27$ \citep{fuhr1988}, $A_{\mathrm{Fe}}=2.7 \times 10^{-5}$
\citep{wilms00}, $A_{26}=0.5$, $A_{\mathrm{(1s)}}=1.0$, and $T=10^{6}$~K, we obtain
$\tau_{0} = 1$ for $N_{\mathrm{H}}=1.7 \times 10^{21}$~cm$^{-2}$. Many systems have
larger $N_{\mathrm{H}}$ values. The situation approaches to case B or D, which are
manifested by a higher Lyman series photon ($n \rightarrow 1$; $n>2$) degrading into a
Balmer series ($n \rightarrow 2$; $n>2$) photon and a Ly$\alpha$ ($n=2 \rightarrow 1$)
photon or by two photon decay after repeating numerous resonance scattering due to their
large optical depth \citep{menzel1937}.

In general, RT effects depend on the geometry, density and velocity structures, and the
spectral shape of the incident emission. Generic tools widely used in X-ray spectroscopy
are not universally applicable to all systems. Calculations specific to each system are
necessary and should be the norm in the X-ray microcalorimeter era that has just
begun. Also, in practice, numerous lines appear both in emission and absorption,
blending each other and being shifted by different redshifts, in high-resolution X-ray
spectra as we will show in this paper. Synthesized spectra with RT calculations are
needed to disentangle such complexities. This is clearly demonstrated in the X-ray
microcalorimeter data of Cyg X-3 \citep{collaboration2024}.

\medskip

The purpose of this paper is to showcase the utility of the RT calculation in
interpreting the microcalorimeter data, focusing on line ratios, and thereby
constraining the physics of the mass outflow from compact objects. We use a low-mass
X-ray binary Circinus X-1 (Cir X-1), which was observed with XRISM showing particularly
rich spectral features, thus is challenging for modeling. We use the \texttt{cloudy}
code for the RT calculation. We start with a brief description of the source and why we
chose it for our study in \S\ref{s2}. We present the observation and data reduction in
\S\ref{s3}. We give a phenomenological description of the spectrum in \S\ref{s4}. We
conduct the RT calculation and perform line ratio diagnostics as well as spectral
modeling in \S~\ref{s5}. Based on these, we discuss the physical parameters and the
driving mechanism of the outflow in \S\ref{s6}. The conclusion is given in \S\ref{s7}.

\section{Target}\label{s2}
Cir X-1 was discovered at the dawn of X-ray astronomy using an Aerobee 150 rocket
\citep{margon1971}. Since then, the source has been monitored in the X-rays for half a
century, revealing its wild behavior in the flux that varies between a few milli and a
super crab \citep{parkinson2003}, which is unseen in any other X-ray binaries. There has
been long debates on the nature of the enigmatic source, but accumulating evidence
favors that it is a low-mass X-ray binary (LMXB) hosting a NS seen close to an edge-on
view with an inclination angle of 60--75 degree \citep{tominaga2024}.

\begin{figure}[!hbtp]
 \begin{center}
  \includegraphics[width=1.0\columnwidth,bb=0 0 432 288,clip]{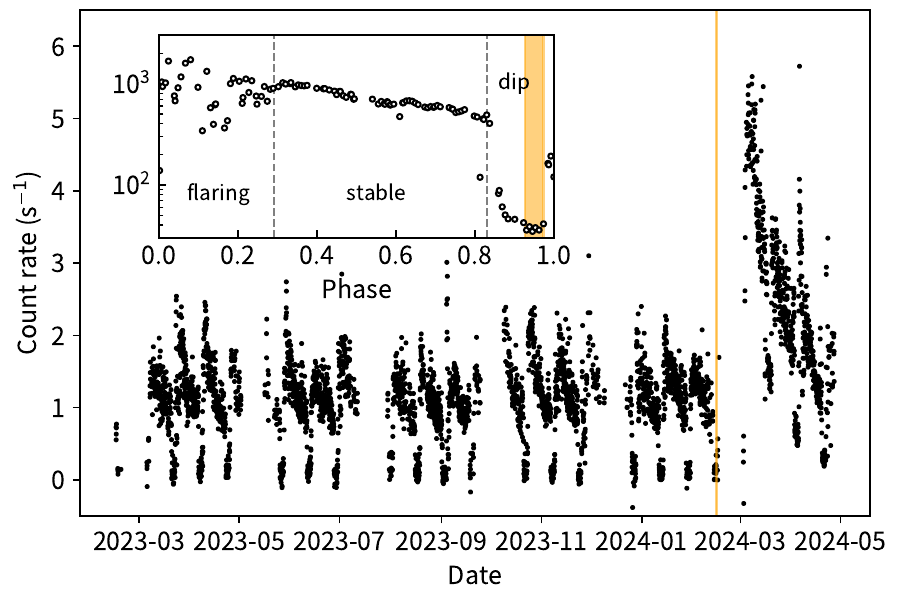}
 \end{center}
 \caption{X-ray light curve in the 2--20~keV band using the Gas Slit Cameras
 \citep{mihara2011} onboard MAXI \citep{matsuoka2009} for one year before the XRISM
 observation (orange) and a few months thereafter. The inset is the phase-folded light
 curve of the NICER observation covering an entire orbital phase in August 2020
 \citep{tominaga2023}. Two cycles after the XRISM observation, the flux increased,
 reaching the highest level since 2014. {Alt text: two scatter plots of X-ray light
 curve of MAXI and NICER.}}
 \label{f02}
\end{figure}

\begin{figure*}[!hbtp]
 \begin{center}
  \includegraphics[width=1.0\textwidth,clip]{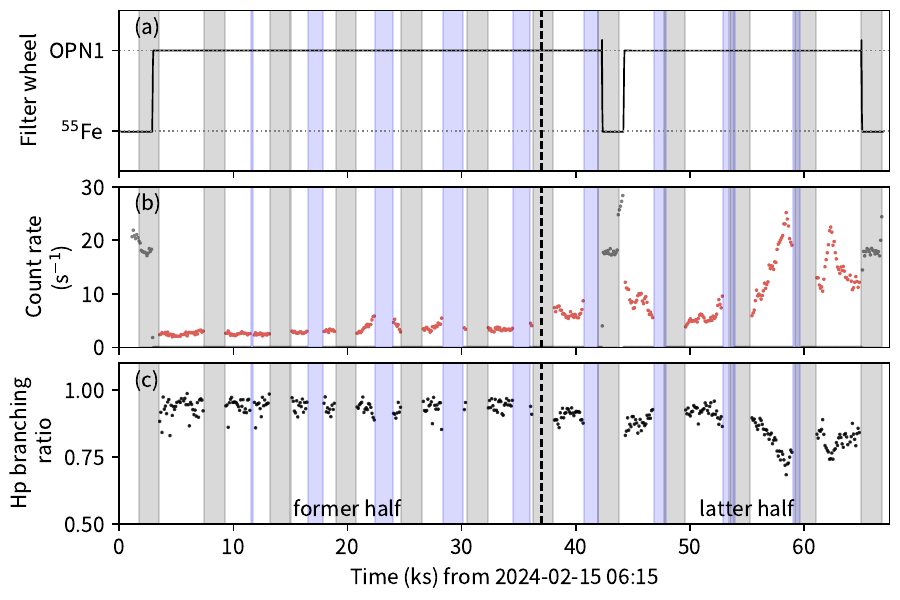}
 \end{center}
 \caption{(a) Filter wheel positions, (b) 2--12~keV count rate binned at 100~s, and (c) the
 fraction of Hp events. The count rate increased during the FW illumination (gray dots),
 which was not used except for gain tracking purposes. The observation was interrupted
 by South Atlantic Anomaly passages (blue) and the occultations by the Earth (gray). The
 observation is divided into two halves (18.1 and 16.8~ks each) by the broken line for
 time-sliced spectroscopy. {Alt text: three scatter plots of filter wheel position,
 X-ray count rate, and the high primary branching ratio.}}
 \label{f21}
\end{figure*}

The binary has an orbital period of 16.6 day \citep{kaluzienski1976} with a large
eccentricity of $\sim$0.45 \citep{Jonker2007a} at a distance of 9.4~kpc
\citep{Heinz2015}. This is one of a few known compact objects in a binary system
associated with a supernova remnant that gave birth to the object, suggesting an extreme
youth of $\sim 5 \times 10^3$~years \citep{Heinz2013}. Despite its youth, the NS is
considered to have a weak magnetic field, which is evidenced by the lack of coherent
X-ray signals, typical behaviors as a Z source \citep{shirey1998}, the presence of twin
kHz quasi-periodic oscillations \citep{Boutloukos2006}, and type I bursts
\citep{tennant1986}, all of which are observational signatures of LMXBs hosting a NS of
a weak magnetic field \citep{vanderklis2006}.

The source was observed with the High-Energy Transition Grating (HETG) spectrometer
\citep{canizares2005} onboard the Chandra X-ray Observatory \citep{weisskopf2000} in the
first year of launch when Cir X-1 was historically bright, exceeding a crab. The X-ray
spectrum taken at the orbital phase $\phi=0.99$ was rich in spectral features of
highly-ionized Ne, Mg, Si, S, and Fe with a P Cygni profile of a velocity dispersion of
$2 \times 10^{3}$~km~s$^{-1}$ \citep{Brandt2000,Schulz2002}. This was the first clear
detection of the P Cygni profile in X-rays, which constitutes strong evidence for the
presence of outflow. This was interpreted as outflows from the accretion disk
\citep{Brandt2000} and opens an avenue for constraining the physical parameters of the
outflowing matter from compact objects through X-ray spectroscopy. However, HETG was
sensitive to the P Cygni profile of the highest end of the velocity dispersion exceeding
$\mathcal{O}$(10$^{3}$~km~s$^{-1}$) if we use highly-ionized Fe features, and the rest
was left for microcalorimeter spectrometers.

The X-ray spectrum of Cir X-1 changes drastically along the orbital phase. During a
decade-long low flux state in 2010's, the change was non-repeatable
\citep{Asai2014a}. Since around July 2019, the flux has recovered to a few tenths of a
crab and the orbital change has become repeatable, which continued until the XRISM
observation (figure~\ref{f02}). \citet{tominaga2023} conducted X-ray observations with
the Neutron Star Interior Composition Explorer (NICER; \citealt{Gendreau2016}) that
covered the entire orbital phase at high cadence in August 2020. They confirmed three
phases based on flux variation (figure~\ref{f02} inset) proposed earlier \citep{iaria2001a}; the
dip phase ($\phi=0.84-1.0$) when the flux decreases to a few percent, the flaring phase
($\phi=0.0-0.29$) when the flux exhibits rapid variability, and the stable phase
($\phi=0.29-0.83$) when the flux gradually declines. X-ray spectra were quite different
among the phases or even within the same phase, but they were explained by the same
phenomenological model with different parameters. The major factor for the variability
was found to be the column density and the partial covering fraction of the absorber in
the line of sight \citep{brandt1996}. Emission and absorption features from the
photoionized plasma were also found. The line ratios were stable over the phase despite
the apparent change in the continuum flux. Based on this, \citet{tominaga2023} came to a
picture shown in their figure~6, based on which we will develop our interpretation of
the XIRSM data to be shown below.

\begin{figure*}[!hbtp]
 \begin{center}
 \includegraphics[width=0.48\textwidth,clip]{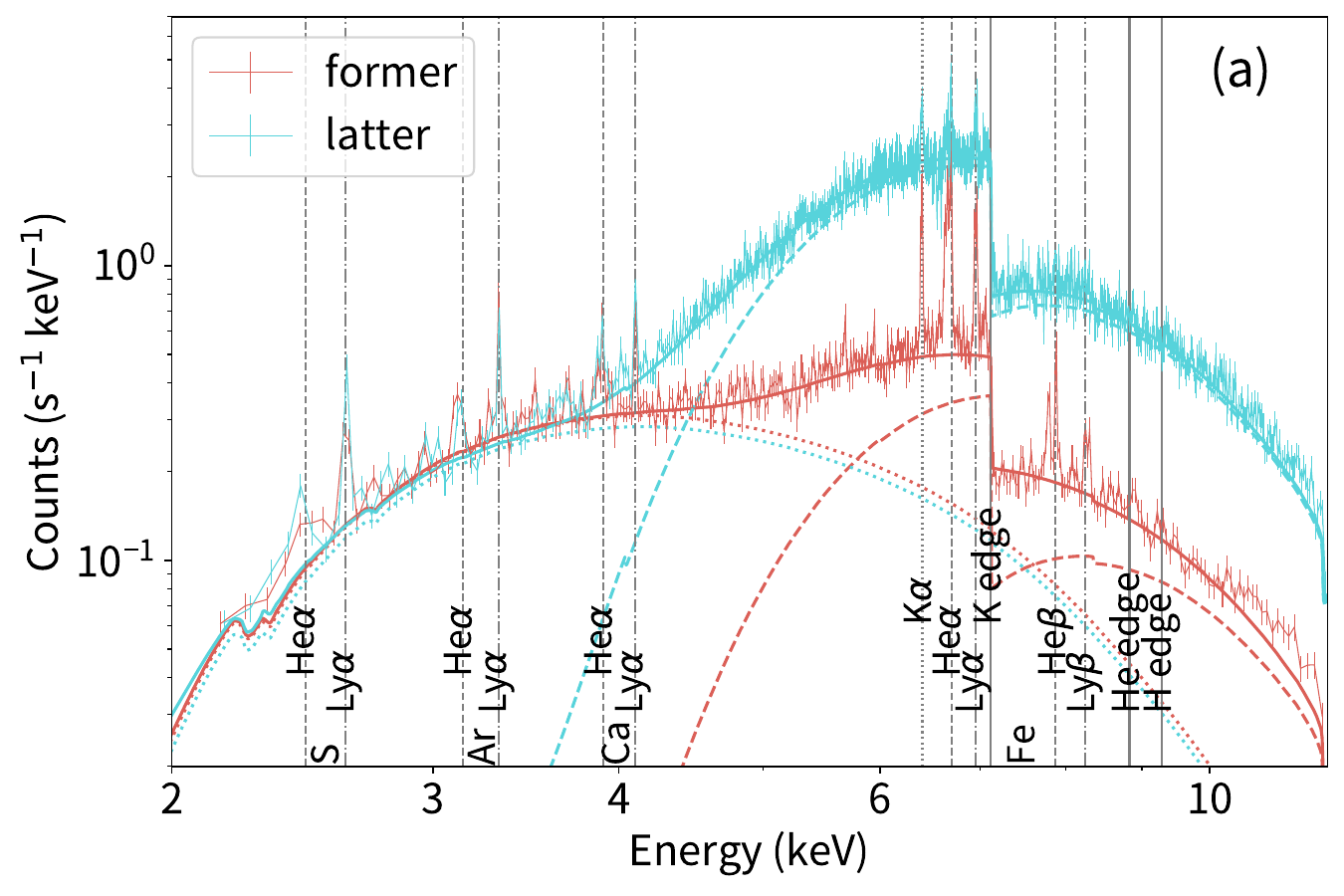}
 \includegraphics[width=0.48\textwidth,clip]{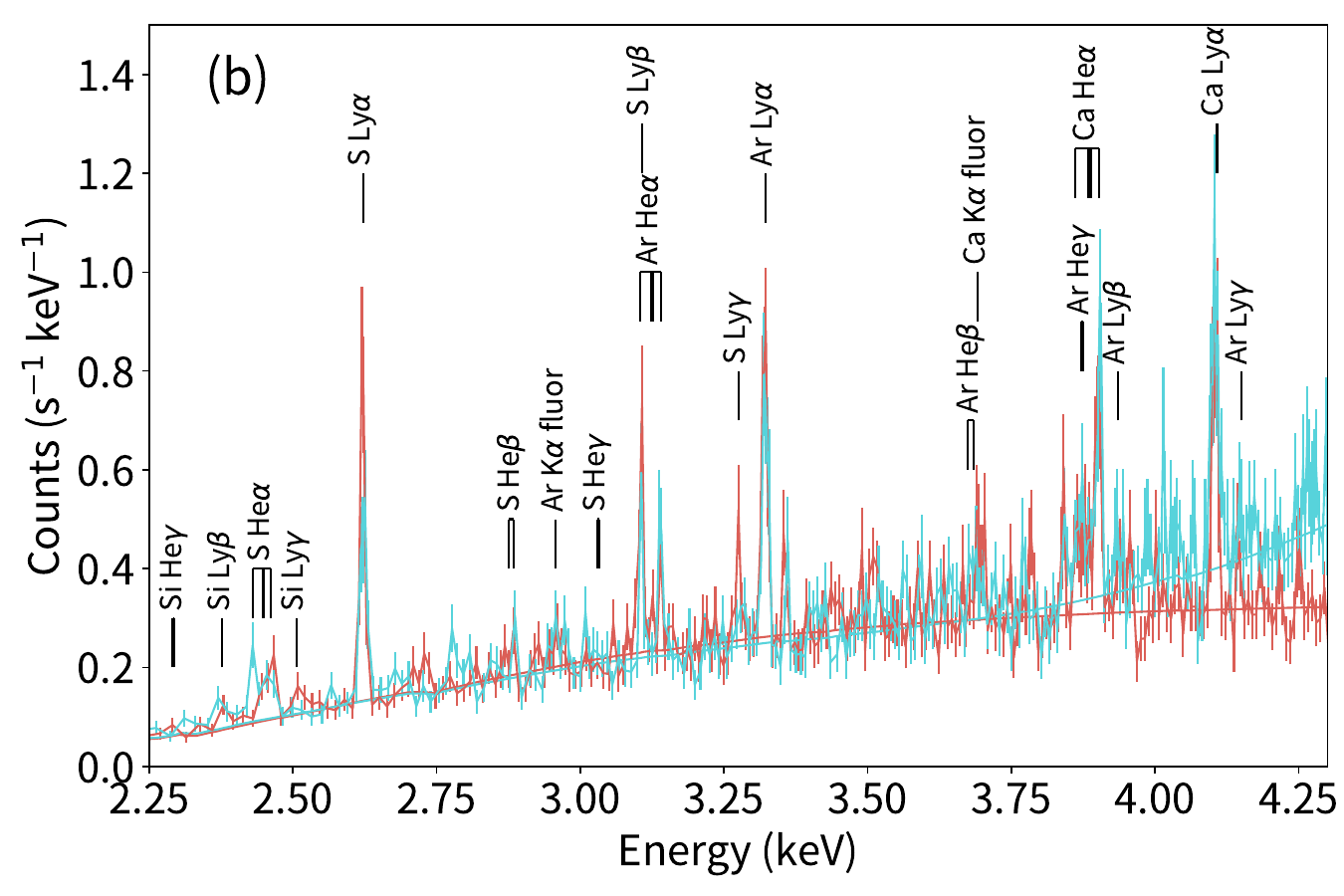}
 \includegraphics[width=0.48\textwidth,clip]{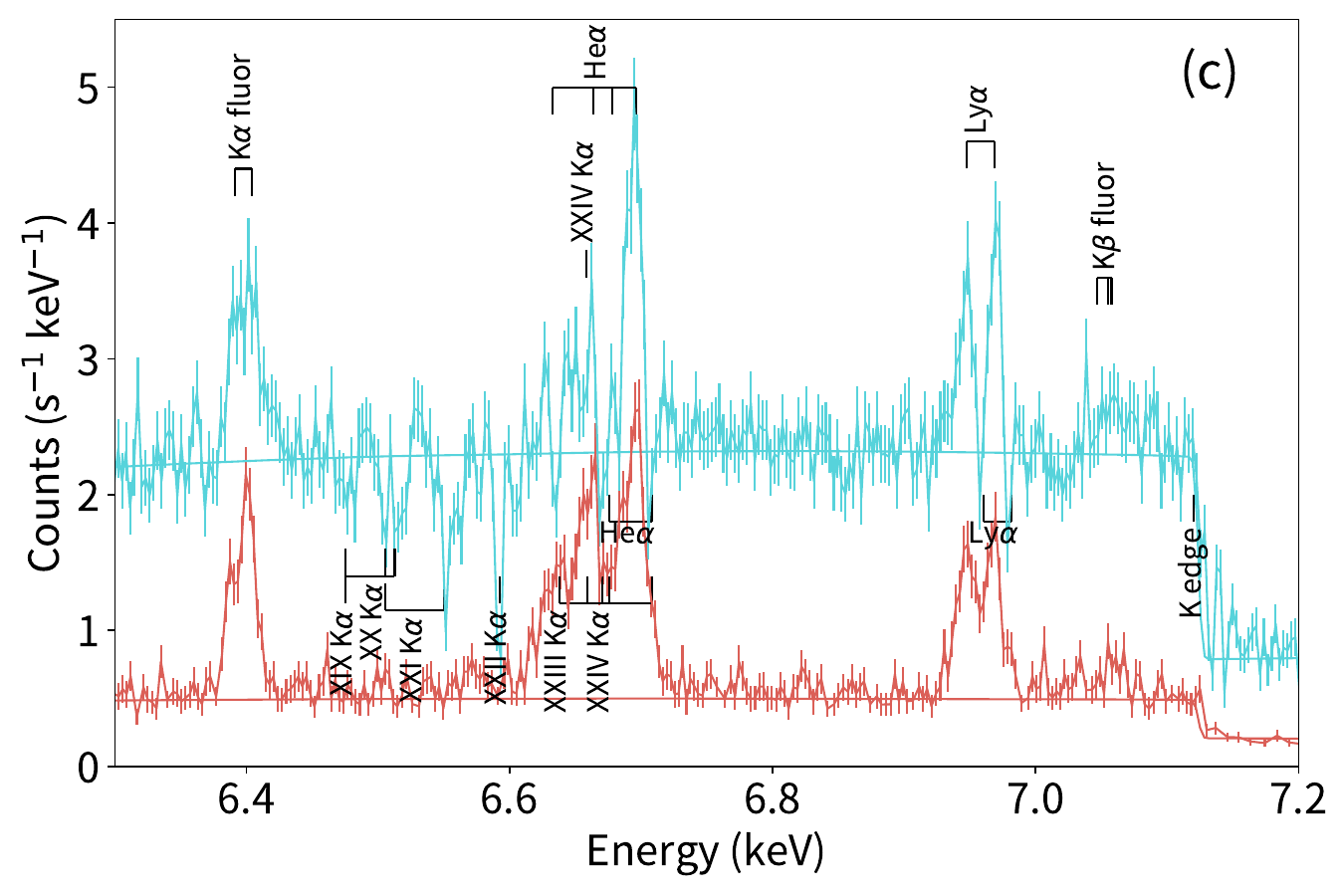}
 \includegraphics[width=0.48\textwidth,clip]{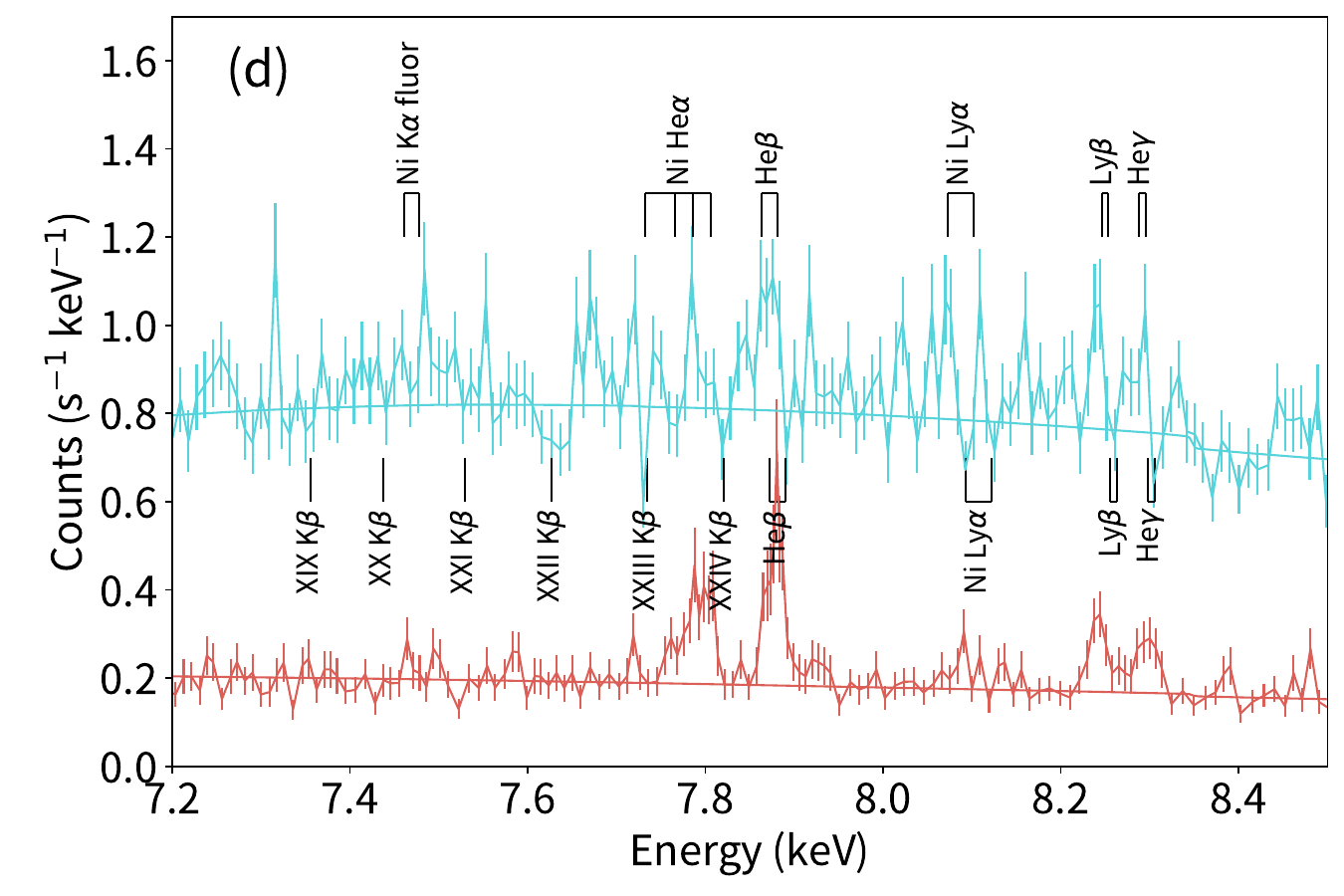}
 \end{center}
 \caption{(a) Total (2--12~keV), (b) soft (2.25--4.25~kev), (c) Fe K$\alpha$
 (6.3--7.2~keV), and (d) Fe K$\beta$ (7.2--8.5~keV) band spectra of the former (red) and
 latter (cyan) half of the observation. The absorption lines are multiplicative and the
 emission lines are additive upon the continuum emission, thus their variations can be
 better viewed respectively in the logarithmic (a) and the linear (b--d) 
 scale. The best-fit continuum model is shown: total (solid), absorbed (dotted) and
 unabsorbed (dashed) by the partial covering absorber.
 (b--d) Emission line complexes are shown on the upper side, while absorption line
 complexes are on the lower side of the data. Lines without an element name are Fe. The
 line labels are shifted by representative velocities for clarity: $+$750~km~s$^{-1}$ for Ni
 Ly$\alpha$ $+$360~km~s$^{-1}$ for the other Fe and Ni emission lines except for the
 fluorescence line, null for the other emission lines, and $-$180~km~s$^{-1}$ for all
 the absorption lines. {Alt text: four line plots of X-ray spectra in different energy
 bands.}}
 \label{f06}
\end{figure*}

\section{Observations and Data Reduction}\label{s3}
\subsection{Observation}\label{s3-1}
The observation (sequence number 300028010) was made from 2024-02-15 06:19 to 2024-02-16
00:51 as a performance verification program. The duration covers an orbital phase of
$\phi=$0.926--0.973 in the ephemeris by \citet{nicolson2007}. We targeted just before
the phase origin $\phi=0.0$ in the last part of the dip phase (figure~\ref{f02} inset),
which interested us the most among the NICER data covering an entire orbit
\citep{tominaga2023}.

In this paper, we focus on the \textit{Resolve} data. The bandpass is limited to above
$\sim$1.7~keV due to the cryostat window yet to be opened
\citep{midooka2021a}. Figure~\ref{f21} shows the time series of some selected house
keeping and X-ray event data. During the $\sim$70~ks telescope time, the observation was
interrupted by South Atlantic Anomaly (SAA) passages and occultation by the Earth. No
recycling operation of the adiabatic demagetization refrigerator was performed
\citep{shirron2024}. At the beginning, middle, and end of the observations, the filter
wheel \citep{shipman2024} was rotated to illuminate the microcalorimeter pixels with the
$^{55}$Fe sources for gain tracking calibrations \citep{porter16}. Otherwise, the filter
wheel was at an open position. The total on-source integration is 33.7~ks.

\subsection{Data reduction}\label{s3-2}
We started with the level 2 products of the pipeline processing \citep{doyle2022}
version 03.00.011.008. The standard processing is adequate for this source in the right
dynamic range of the count rate of $\mathcal{O}$(10 s$^{-1}$); the background is
negligible \citep{kilbourne2018} and artifacts due to high count rates are minimum
\citep{mizumoto2022}. We selected Hp grade events. Here, the Hp grade is given for
events without any other events overlapping in time in the same pixel, which is well
calibrated for energy as of writing. We also removed events below 0.3~keV, those with
too fast or slow pulse rise times for their energy \citep{mochizuki2024}, and those
recorded with pixel 27, which is known to behave unpredictably in gain variation. A
total of 0.2 million events were left.

During the observation, the count rate varied (figure~\ref{f02}b) and so did the Hp
grade branching ratio as a result (figure~\ref{f02}c). Their average values are
6.4~s$^{-1}$ and 0.86. The detector and telescope response files were generated for the
observation duration of interest when spliced respectively using the \texttt{rslmkrmf}
and \texttt{xaarfgen} tools in the \texttt{HEASoft} package version 6.34. The background
spectrum was not subtracted. Pixel-to-pixel spectral variation of unidentified reasons
is known, which may affect the global fitting results, but not the results based on line
ratios. The errors quoted hereafter represent a statistical uncertainty of 1 $\sigma$.

\begin{figure*}[!hbtp]
 \begin{center}
  \includegraphics[width=0.45\textwidth,bb=100 160 588 540, clip]{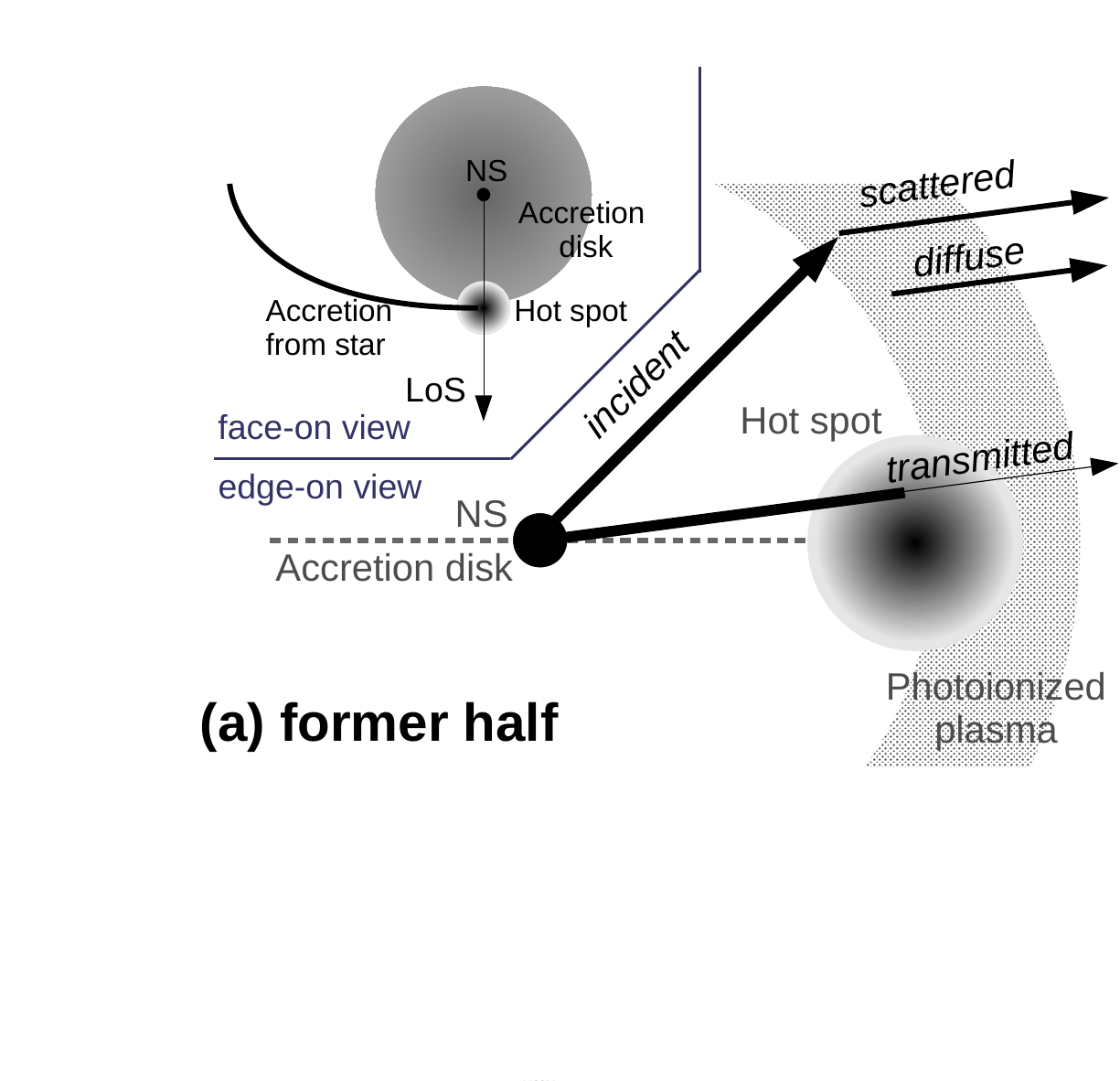}
  \includegraphics[width=0.45\textwidth,bb=100 160 588 540, clip]{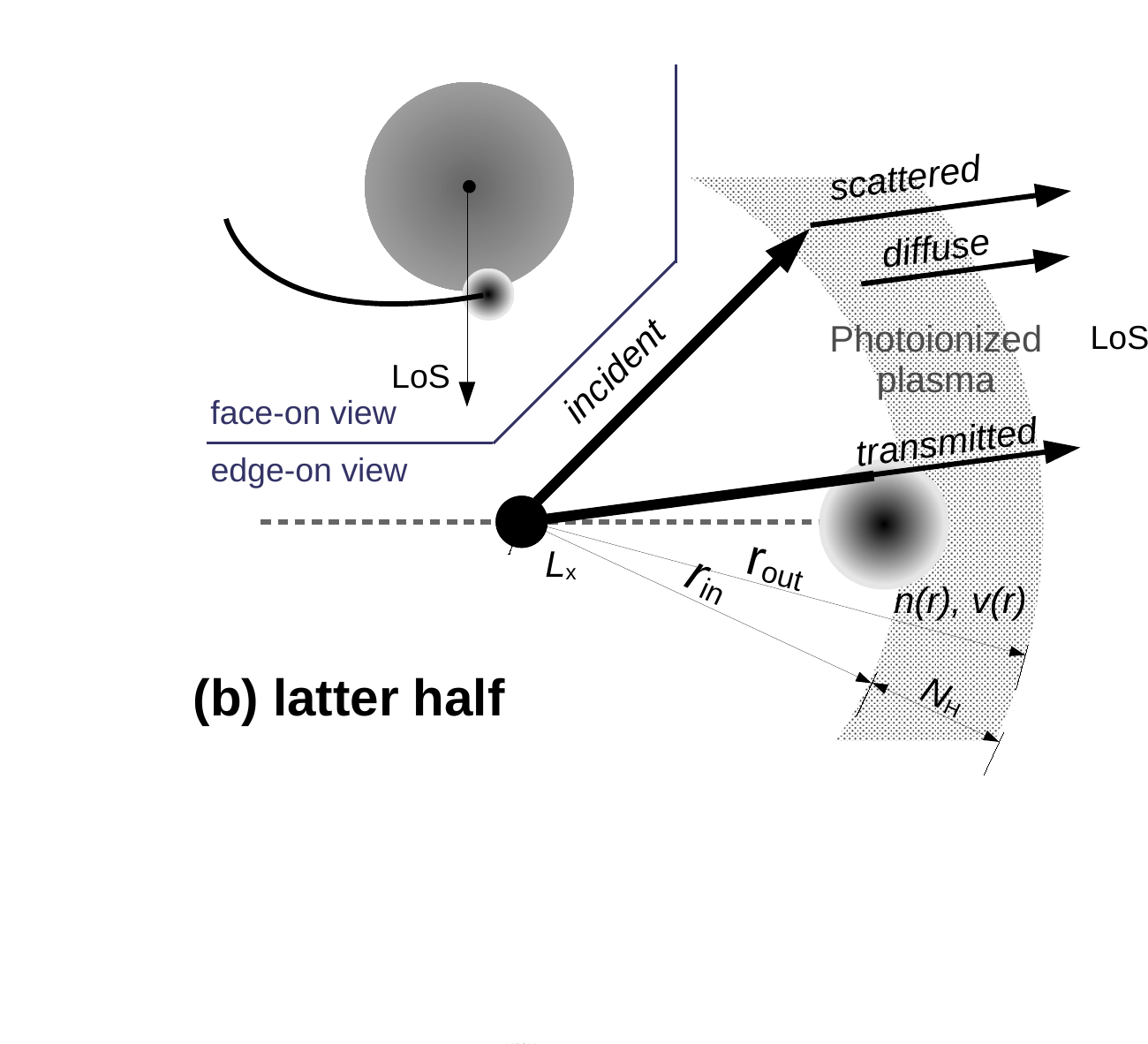}
 \end{center}
 \caption{Schematic view of the system for the photoionized plasma. The face-on (inset)
 and edge-on (main) views of the (a) former and (b) latter half of the observation are
 shown. The main difference is the clocking of the hot spot in the face-on view as the
 orbital phase progresses. This makes the change in the blocking fraction of the
 incident emission by the hot spot in the line of sight (LoS). The physical parameters
 used in the RT modeling is shown in (b). {Alt text: two schematic views of the system
 for the former and the latter half of the observation.}}
 \label{f00}
\end{figure*}

\section{Analysis}\label{s4}
The goal of this section is to describe the spectrum using phenomenological models. We
start by visually inspecting the data to identify spectral features and components
(\S\ref{s4-1}) and describe each component (\S\ref{s4-2}) separately for the continuum
(\S\ref{s4-2-1}), emission lines (\S\ref{s4-2-2}), and absorption lines
(\S\ref{s4-2-3}).

\subsection{Inspection}\label{s4-1}
The X-ray count rate is variable by an order of magnitude during the observation
(figure~\ref{f21}b). The behavior is different in the former and the latter half. The
X-ray spectrum is also different between the two halves. Figure~\ref{f06} (a) shows the
2--12~keV band spectrum for the two halves. At least, two spectral components are
recognized; the soft component is dominant below $\sim$4~keV and is non-variable, while
the hard component is dominant above $\sim$4~keV and is variable.

The soft component (figure~\ref{f06}b) is characterized by emission lines of
highly-ionized S, Ar, and Ca of $n = \{2, 3, 4\} \rightarrow 1$ transitions, which we
respectively denote as Ly$\alpha$, Ly$\beta$, and Ly$\gamma$ for the H-like and
He$\alpha$, He$\beta$, and He$\gamma$ for the He-like ions. The same component may
account for the emission lines of highly-ionized Fe above $\sim$4~keV for the $n = 2
\rightarrow 1$ transitions (figure~\ref{f06}c) and the $n = \{3, 4\} \rightarrow 1$
transitions (figure~\ref{f06}d) as well as the emission lines of highly-ionized Ni for
the $n = 2\rightarrow 1$ transitions (figure~\ref{f06}d).

The hard component suffers strong absorption by neutral matter, which is evident as the
gradual attenuation toward the soft end and the photoionization K edge of Fe at
7.12~keV. The gradual attenuation is likely due to partial coverage by the neutral
matter, as the attenuation toward the soft end is not as rapid as complete blocking. The
depth of the Fe K edge is similar between the former and the latter half in the
logarithmic scale (figure~\ref{f06}a), indicating that the absorbing column did not
change much.

In the latter half of the observation, the hard component also suffers absorption by
mildly-ionized Fe for the K$\alpha$ ($n = 2\rightarrow 1$) and K$\beta$ ($n = 3
\rightarrow 1$) transitions respectively in the 6.4--6.7 keV (figure~\ref{f06}c) and
7.3--7.9 keV (figure~\ref{f06}d) band. These are caused by the inner-shell excitation
absorption from the ground state to excited levels of mildly ionized ions. The
photoelectric absorption cross section exhibits strong resonance structures at these
energies \citep{behar2002,bautista2004,kallman2004}. Some hints of such absorption
features were obtained in the HETG observations \citep{Schulz2008} for the K$\alpha$
lines, which were clearly identified with \textit{Resolve}. The new detection of the
K$\beta$ absorption lines is a consequence of two unique capabilities of
\textit{Resolve} besides the energy resolution: (i) its high quantum efficiency beyond
8~keV due to the HgTe X-ray absorber as opposed to the Si absorber in the conventional
X-ray spectrometers and (ii) a very low background equivalent to only one event per
spectral bin per 100~ks that allows detection of continuum emission with a high
signal-to-noise ratio, upon which the absorption features are imprinted.

For the $n = \{2, 3, 4\}\rightarrow 1$ transitions of highly-ionized Fe, the features
appear only in emission in the former half but appear both in emission and absorption in
the latter half (figure~\ref{f06}c and d). The emission features are red-shifted, while
the absorption features are blue-shifted, forming a P Cygni profile. This is
evident in the $w$ line at 6.700~keV in the He$\alpha$ complex and the Ly$\alpha_1$ and
Ly$\alpha_2$ lines respectively at 6.973 and 6.952 keV in the Ly$\alpha$ complex. Here,
we use $^{1}P_{1} \rightarrow ^{1}S_{0}$, $^{3}P_{1,2} \rightarrow ^{1}S_{0}$,
and $^{3}$S$_{1} \rightarrow ^{1}$S$_{0}$ transitions as $w$, $x$, $y$, and $z$ lines in
the He complex and $^{2}P_{3/2} \rightarrow ^{2}S_{1/2}$ and $^{2}P_{1/2}
\rightarrow ^{2}S_{1/2}$ transitions as Ly$_1$ and Ly$_2$ lines in the Ly
complex. These fine-structure levels were resolved with the \textit{Resolve}'s energy
resolution for the first time in space except for the Sun \citep{tanaka1986}.

By continuing the visual inspection, we can further deduce that the emission lines of
the highly ionized Fe do not suffer from the absorption causing the Fe K edge at
7.12~keV. This is suggested from the Lyman decrement of H-like Fe; i.e. the line ratio
between Fe Ly$\beta$ at 8.25~keV over Ly$\alpha$ at 7.0~keV. The Ly$\beta$ line is above
and the Ly$\alpha$ line is below the Fe K edge. The face value of the Lyman decrement is
$\sim$0.2 in the former half of the spectrum. If we correct for the absorption
represented by the Fe K edge, assuming that the Ly$\beta$ emission line was attenuated
by it, the corrected Lyman decrement would be $\sim$1.0, which cannot be achieved under
any plasma conditions as we show later in \S\ref{s5-1}.

\subsection{Description}\label{s4-2}
\subsubsection{Continuum}\label{s4-2-1}
For the phenomenological description of the spectra, we start with the continuum
emission. We follow the approach in \citet{tominaga2023}, in which the broad-band
spectrum is described by the partially-covered multi-color disk blackbody emission. The
simple model successfully explained the varying spectra taken over an entire orbital
phase with NICER.

We applied the model both for the former and the latter spectra in energy ranges devoid
of spectral lines. The fitting parameters are the innermost temperature
($T_{\mathrm{in}}$) and the normalization ($N^{\mathrm{(disk)}}$) of the disk blackbody
emission, the H-equivalent column density ($N_{\mathrm{H}}^{\mathrm{(pca)}}$), the
covering fraction ($f^{\mathrm{(pca)}}$), and the Fe abundance
($A_{\mathrm{Fe}}^{\mathrm{(pca)}}$) of the partial covering absorber (PCA). The Fe
abundance parameter is intended to adjust for the Fe K edge depth. The entire spectrum
is attenuated by the interstellar matter (ISM) of the H-equivalent column density
($N_{\mathrm{H}}^{\mathrm{(ism)}}$). We tied the $N_{\mathrm{H}}^{\mathrm{(ism)}}$,
$A_{\mathrm{Fe}}^{\mathrm{(pca)}}$, and $T_{\mathrm{in}}$ values common and others
independent between the two spectra.  This resulted in a reasonable fit with a
reduced $\chi^{2}$/d.o.f $=$1.04/2483. The best-fit parameters are shown in
table~\ref{t03} and the model in figure~\ref{f06}.

\begin{table}[!hbtp]
 \tbl{Best-fit parameters for the continuum emission.}{
 \begin{tabular}{lllcc}
  \hline
  Comp & Par & Unit & Former & Latter \\
  \hline
  Disk\footnotemark[$*$] & $T_{\mathrm{in}}$ & (keV) & \multicolumn{2}{c}{$2.12^{+0.04}_{-0.04}$} \\
  & $N^{\mathrm{(disk)}}$ & & $3.88^{+0.52}_{-0.46}$ & $16.0^{+1.7}_{-1.6}$\\
  PCA\footnotemark[$\dagger$] & $N_{\mathrm{H}}^{\mathrm{(pca)}}$ & ($10^{22}$
	  cm$^{-2}$) & $95.9^{+4.3}_{-4.2}$ & $74.1^{+1.4}_{-1.3}$\\
  & $f^{\mathrm{(pca)}}$ & & $0.91^{+0.01}_{-0.01}$ & $0.979^{+0.001}_{-0.001}$\\
  & $A_{\mathrm{Fe}}^{\mathrm{(pca)}}$ & & \multicolumn{2}{c}{$0.99^{+0.03}_{-0.03}$} \\
  ISM\footnotemark[$\dagger$] & $N_{\mathrm{H}}^{\mathrm{(ism)}}$ & (10$^{22}$ cm$^{-2}$) & \multicolumn{2}{c}{$0.98^{+0.25}_{-0.24}$} \\
  \hline
 \end{tabular}}
 \label{t03}
 \begin{tabnote} 
  \footnotemark[$*$] The \texttt{diskbb} model \citep{mitsuda1984} of \texttt{xspec} is
  used. The normalization is scaled as $N^{\mathrm{(disk)}} \equiv
  (R_{\mathrm{in}}/D_{10})^{2} \cos{\theta}$, in which $R_{\mathrm{in}}$ is the inner
  disk radius (km), $D_{10}$ is the distance in the unit of 10~kpc, and $\theta$ is the
  viewing angle of the disk.\\
  \footnotemark[$\dagger$] The \texttt{tbpcf} model for PCA and \texttt{tbabs} model
  for ISM \citep{wilms00} is used.\\
 \end{tabnote} 
\end{table}

As expected, the absorbed and non-absorbed components by the PCA respectively account
for the continuum emission below and above $\sim$4~keV. The spectral change above 4~keV
is mostly caused by the increase of $N^{\mathrm{(disk)}}$ by $\sim$4 times, while the
other parameters change much less of $\lesssim$20\%. The flux of the non-absorbed
component can be assessed with $N^{\mathrm{(disk)}} (1-f^{\mathrm{(pca)}})$, which
remained the same within the error between the two spectra despite their independence in
the fitting. The Fe K edge at 7.12~keV is fully explained by the PCA for
$A_{\mathrm{Fe}}^{\mathrm{(pca)}}$ being consistent with unity.

\subsubsection{Emission lines}\label{s4-2-2}
We derive phenomenological parameters of emission line features using the former half of
the spectrum, which is less contaminated by absorption features. We focus on conspicuous
lines, including the Ly$\alpha$ line complex of S, Ar, Ca, Fe, and Ni and the
Ly$\beta$ complex of Fe for the H-like ions and the He$\alpha$ line complex of S, Ar,
Ca, Fe, and Ni and the He$\beta$ complex of Fe for the He-like ions. We also
include the Fe K$\alpha$ fluorescence line complex.

For the Ly$\alpha$ and Ly$\beta$ complexes, we applied two Gaussian models representing
the Ly$_{1}$ and Ly$_{2}$ lines plus the power-law model representing the local
underlying continuum. The intensity ratio of the two lines was fixed to 2:1. Their
normalization, width, and redshift were fitted collectively for each complex. The
exception is the Fe Ly$\alpha$ line, which has sufficient statistics to determine line
normalization and width independently.
For the He$\alpha$ and He$\beta$ complexes, we used four Gaussian models representing
the $w$, $x$, $y$, and $z$ lines plus the power-law model. The intensity of the lines
was fitted individually, while the redshift and the line width were fitted collectively
for each complex.
For the Fe K$\alpha$ fluorescence line, we used two Gaussian models instead of a more
elaborate model \citep{holzer1997} for simplicity. Table~\ref{t01} shows the result.

\begin{table} 
 \tbl{Phenomenological model fitting for the emission lines in the former half.}{
 \begin{tabular}{lcccc}
  \hline
  Label & $E$\footnotemark[$*$] & $\log{f}$\footnotemark[$\dagger$] & $I$\footnotemark[$\ddagger$] & $v$\footnotemark[\S] \\
        & (keV) & & (ks$^{-1}$~cm$^{-2}$) & (km~s$^{-1}$) \\
  \hline
  S \emissiontype{XV} He$\alpha$ ($w$) & 2.461 & --0.12 & $0.07 \pm 0.02$ & $-336 \pm 70$\\
  S \emissiontype{XVI} Ly$\alpha_1$ & 2.623 & --0.56 & $0.14 \pm 0.02$ & $75 \pm 49$\\
  \hline
  Ar \emissiontype{XVII} He$\alpha$ ($w$) & 3.140 & --0.12 & $0.02 \pm 0.01$ & $-179 \pm 41$\\
  Ar \emissiontype{XVI} Ly$\alpha_1$ & 3.323 & --0.56 & $0.06 \pm 0.01$ & $49 \pm 59$\\
  \hline
  Ca \emissiontype{XIX} He$\alpha$ ($w$) & 3.902 & --0.12 & $0.04 \pm 0.01$ & $103 \pm 83$\\
  Ca \emissiontype{XX} Ly$\alpha_1$ & 4.108 & --0.56 & $0.04 \pm 0.01$ & $26 \pm 45$\\
  \hline
  Fe fluor K$\alpha_1$ & 6.404 & ... & $0.12^{+0.01}_{-0.01}$ & $75 \pm 27$\\
  Fe fluor K$\alpha_2$ & 6.391 & ... & $0.07^{+0.01}_{-0.01}$ & (tied w. above)\\
  \hline
  Fe \emissiontype{XXV} He$\alpha$ ($w$) & 6.700 & --0.14 & $0.28 \pm 0.01$ & $282 \pm 23$\\
  Fe \emissiontype{XXV} He$\alpha$ ($x$) & 6.682 & --4.80 & $0.00 \pm 0.00$ & (tied w. above)\\
  Fe \emissiontype{XXV} He$\alpha$ ($y$) & 6.668 & --1.20 & $0.22 \pm 0.01$ & (tied w. above)\\
  Fe \emissiontype{XXV} He$\alpha$ ($z$) & 6.637 & --6.50 & $0.12 \pm 0.01$ & (tied w. above)\\
  Fe \emissiontype{XXV} He$\beta$ ($w$) & 7.881 & --0.86 & $0.03 \pm 0.01$ & (unconstrained)\\
  Fe \emissiontype{XXVI} Ly$\alpha_1$& 6.973 & --0.56 & $0.12 \pm 0.01$ & $237 \pm 26$\\
  Fe \emissiontype{XXVI} Ly$\alpha_2$& 6.952 & --0.87 & $0.11 \pm 0.01$ & (tied w. above)\\
  Fe \emissiontype{XXVI} Ly$\beta_1$ & 8.253 & --1.30 & $0.02 \pm 0.00$ & $299 \pm 74$\\
  \hline
  Ni \emissiontype{XXVII} He$\alpha$ ($w$) & 7.806 & --0.15 & $0.03 \pm 0.01$ & $90 \pm 76$\\
  Ni \emissiontype{XXVIII} Ly$\alpha_1$ & 8.102 & --0.56 & $0.01 \pm 0.00$ & $447 \pm 276$\\
  \hline
 \end{tabular}}
 \label{t01} 
 \begin{tabnote} 
 \footnotemark[$*$] Energy from the \texttt{AtomDB} database \citep{smith01,foster2020}.\\
 \footnotemark[$\dagger$] Logarithm of the oscillator strength from the upper to lower
  level for highly-ionized ions from the \texttt{AtomDB} database
  \citep{smith01,foster2020}.\\
 \footnotemark[$\ddagger$] Line intensity from the fitting.\\
 \footnotemark[\S] Energy shift from the fitting. The redward shift is positive.\\
 \end{tabnote} 
\end{table}

\subsubsection{Absorption lines}\label{s4-2-3}
We describe the absorption features using the latter half of the observation. We see
absorption lines of K$\alpha$ ($n=2 \rightarrow 1$) transitions of Fe \emissiontype{XIX}--\emissiontype{XXVI} in the
6.45--7.05~keV (figure~\ref{f06}c) band and their K$\beta$ ($n=3 \rightarrow 1$)
counterparts in the 7.2--8.2~keV band (figure~\ref{f06}d). The lines except for Fe \emissiontype{XXV}
and Fe \emissiontype{XXVI} are formed by the inner-shell excitation transitions from the ground state
to an excited state. There are numerous allowed transitions between them, but only
a fraction is registered in the current atomic databases. We refer to \texttt{chianti}
version 10.1 \citep{delzanna2021}, which has more entries than \texttt{AtomDB} version
3.0.9 \citep{smith01,foster2020} for the K$\alpha$ transitions. None is available for
the K$\beta$ transitions in both. We further restricted ourselves to using lines with an
experimentally verified wavelength for a reliable energy shift determination.

Upon the best-fit continuum (\S~\ref{s4-2-1}) and the emission line (\S~\ref{s4-2-2})
models, we multiplied a Gaussian absorption line for the transitions expected from the
RT simulation described later (\S~\ref{s5-1-2}). We derived the best-fit optical depth
individually for each line except for the Ly$\alpha$ doublet fixed to a 2:1 ratio. We
derived the energy shift collectively as $-249_{-16}^{+16}$~km~s$^{-1}$ for Fe
\emissiontype{XXV} and \emissiontype{XXVI} and $-343_{-10}^{+10}$~km~s$^{-1}$ for the
remainder.  The result is given in table~\ref{t02}.

\begin{table} 
 \tbl{Phenomenological model fitting for the absorption lines in the latter half.}{
 \begin{tabular}{lcllcc}
  \hline
  Label\footnotemark[$*$] & $E$\footnotemark[$\dagger$] & Fe & Upper
	      level\footnotemark[$\ddagger$] & $B_{r}(k,i)$\footnotemark[$\S$] & $\tau$\footnotemark[$\|$] \\
        & (keV) & & & & \\
  \hline
  \multicolumn{6}{l}{(outer-shell transitions)}\\
  Ly$\alpha_1$ & 6.9732 & \emissiontype{XXVI} & 2p~$^2P_{3/2}$ & 1 & $0.29_{-0.04}^{+0.05}$\\
  Ly$\alpha_2$ & 6.9520 & \emissiontype{XXVI} & 2p~$^2P_{1/2}$ & 1 & (tied w. above)\\
  $w$ & 6.7005 & \emissiontype{XXV} & 1s2p~$^1P_{1}$ & 1 & $0.27_{-0.05}^{+0.05}$\\
  \hline
  \multicolumn{6}{l}{(inner-shell transitions)}\\
  $q$ & 6.6622 & \emissiontype{XXIV} & 1s2s2p~$^2P_{3/2}$ & 1.00 & $0.45_{-0.06}^{+0.07}$\\
  $r$ & 6.6528 & \emissiontype{XXIV} & 1s2s2p~$^2P_{1/2}$ & 0.88 & $0.37_{-0.04}^{+0.05}$\\
  $E1$ & 6.6288 & \emissiontype{XXIII} & 1s2s$^2$2p~$^1P_1$ & 0.75 & $0.04_{-0.04}^{+0.04}$\\
  $B$ & 6.5861 & \emissiontype{XXII} & 1s2s$^2$2p$^2$~$^2P_{1/2}$ & 0.63 & $0.71_{-0.08}^{+0.11}$\\
                     &       &      & 1s2s$^2$2p$^2$~$^2D_{3/2}$ & 0.46 & (blend w. above)\\
  $C1$ & 6.5442 & \emissiontype{XXI} & 1s2s$^2$2p$^3$~$^3D_{1}$ & 0.36 & $0.46_{-0.07}^{+0.07}$\\
  $N2$ & 6.5068 & \emissiontype{XX} & 1s2s$^2$2p$^4$~$^4P_{3/2}$ & 0.29 & $0.16_{-0.05}^{+0.05}$\\
  $N1$ & 6.4970 & \emissiontype{XX} & 1s2s$^2$2p$^4$~$^4P_{5/2}$ & 0.25 & $0.16_{-0.05}^{+0.05}$\\ 
  \hline
 \end{tabular}}
 \label{t02} 
 \begin{tabnote} 
 \footnotemark[$*$] Line labels for the inner-shell transitions follow \citet{rudolph2013}.\\
 \footnotemark[$\dagger$] Energy from \citet{kramida1999} for the outer-shell and 
  \citet{rudolph2013} for the inner-shell transitions.\\
 \footnotemark[$\ddagger$] Upper level of the transition. The lower level is the ground
  state of each ion.\\
 \footnotemark[$\S$] Radiative branching ratio of the transition from the upper ($k$) to
  the lower ($i$) level. Note that $B_{r}(k,i) \equiv
  \frac{A_r(k,i)}{A_r(k)+A_a(k)}$ is different from the fluorescence yield
  $\omega_{r}(k) \equiv \frac{A_r(k)}{A_r(k)+A_a(k)}$, in which $A_{x}(k) = \sum_i
  A_{x}(k,i)$ for the fluorescence ($x=r$) and Auger ($x=a$) decay. The definitions and
  values for the inner-shell transitions are from \citet{palmeri2003b}. \\
 \footnotemark[$\|$] Line optical depth from the fitting.\\
 \end{tabnote} 
\end{table}

\section{Modeling}\label{s5}
We conduct RT calculations (\S~\ref{s5-1}), present line ratio analyses (\S~\ref{s5-2}),
and perform spectral fitting (\S~\ref{s5-3}).

\subsection{Radiative transfer calculation}\label{s5-1}
\subsubsection{Setup}\label{s5-1-1}
We used \texttt{cloudy}, which is a numerical radiative transfer code based on the
one-dimensional, static, and two-stream solver such as \texttt{xstar}
\citep{kallman2001}. It has been used in longer wavelengths in the past, but recent
development has made the code applicable to high-resolution X-ray spectroscopy
\citep{chakraborty2020a,gunasekera2022}. We used a release candidate of the version c25
\citep{gunasekera2024}. For the atomic database, we used the \texttt{chianti} version
10.1 \citep{delzanna2021}.

For the setup (figure~\ref{f00}), we placed a point-like source at the origin for the
incident emission. The emission has an X-ray luminosity ($L_{\mathrm{X}}$) integrated
over 1--1000~Ryd and a spectral shape of the disk blackbody with an innermost
temperature of 2~keV, which best describes the observed continuum emission
(\S\ref{s3-1}). The photoionized plasma is represented by a slab and is characterized by
three parameters: the ionization degree $\xi \equiv L_{\mathrm{X}}/(n
r_{\mathrm{in}}^{2})$ (erg~cm~s$^{-1}$) at the illuminated surface of the slab at a
distance of $r_{\mathrm{in}}$~(cm) from the incident source \citep{tarter1969}, the
density $n$ (cm$^{-3}$) constant over the slab, and the H-equivalent column density
$N_{\mathrm{H}}$ (cm$^{-2}$) across the slab. We assumed no bulk or turbulent
velocities. We ran the simulation with different settings for a grid of the three
parameters logarithmically spaced at 0.1--0.5 dex. We hereafter quote their logarithmic
values without physical units. We also assume that the electron and proton densities are the
same for estimation. 

For each grid, the code produces two spectra (figure~\ref{f00}). One is the transmitted
spectrum through the photoionized plasma, in which absorption features are imprinted
upon the incident continuum emission. The other is the diffuse spectrum from the
photoionized plasma caused by radiative deexcitation and recombination. Scattered
emission also comes from the plasma due to electron scattering, which is not included;
however, we can mimic this component by normalizing the incident emission independent of
the energy by a factor of $e^{-N_{\mathrm{H}}\sigma_{\mathrm{es}}}$, in which
$\sigma_{\mathrm{es}}$ is the electron scattering cross section. We made the
multiplicative absorption model and the additive emission model based on the transmitted
and diffuse spectrum, respectively \citep{porter2006}, so that we can fit the spectra
using the \texttt{xspec} fitting package \citep{arnaud1996}.

\subsubsection{Results}\label{s5-1-2}
\begin{figure}[!hbtp]
 \begin{center}
  \includegraphics[width=1.0\columnwidth,clip]{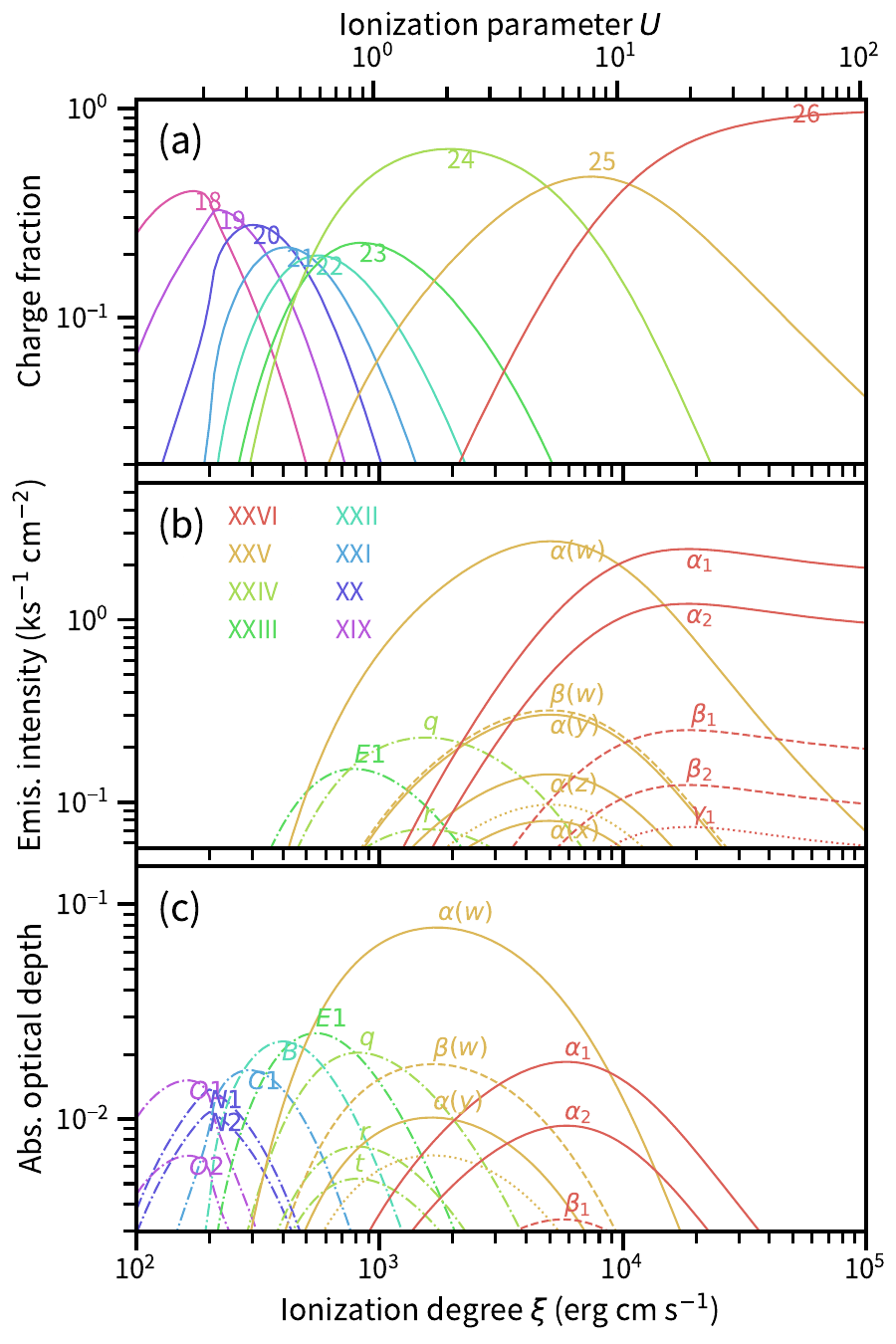}
 \end{center}
 \caption{(a) Charge fraction, (b) emission line intensity, and (c) absorption line
 optical depth of Fe as a function of the ionization degree $\xi$ or the ionization
 parameter $U \equiv Q(H)/4\pi r^{2} n c$, in which $Q(H)$ (s$^{-1}$) is the rate of
 H-ionizing photons calculated with \texttt{cloudy} for
 $N_{\mathrm{H}}=10^{19.5}$~cm$^{-2}$. Emission line intensity is scaled assuming
 $r_{\mathrm{out}}= 1.0 \times 10^{11} (N_{\mathrm{H}}/10^{23.0})^{-0.5}$~cm and the
 distance to the source so as to approximately match with the observed flux. In (a), the
 fractions of the Fe$^{i+}$ ions are shown with an Arabic number $i \in\{18...26\}$,
 where $i=26$ denotes fully-ionized Fe. In (b) and (c), conspicuous lines are shown with
 labels in table~\ref{t02}. Different colors are used for different Fe $j$ feature with
 a Roman number $j \in \{\mathrm{XIX}...\mathrm{XXVI}\}$. Solid, dashed, and dotted
 curves are for $n=4,3,2 \rightarrow 1$ transitions of outer-shell electrons, whereas
 dashed-and-dotted curves are for the K$\alpha$ transitions of inner-shell
 electrons. {Alt text: three line plots of charge state distribution, emission
 intensity, and absorption depth as a function of the ionization degree.}}
 \label{f12}
\end{figure}

\begin{figure*}[!hbtp]
 \begin{center}
 \includegraphics[width=1.0\textwidth,clip]{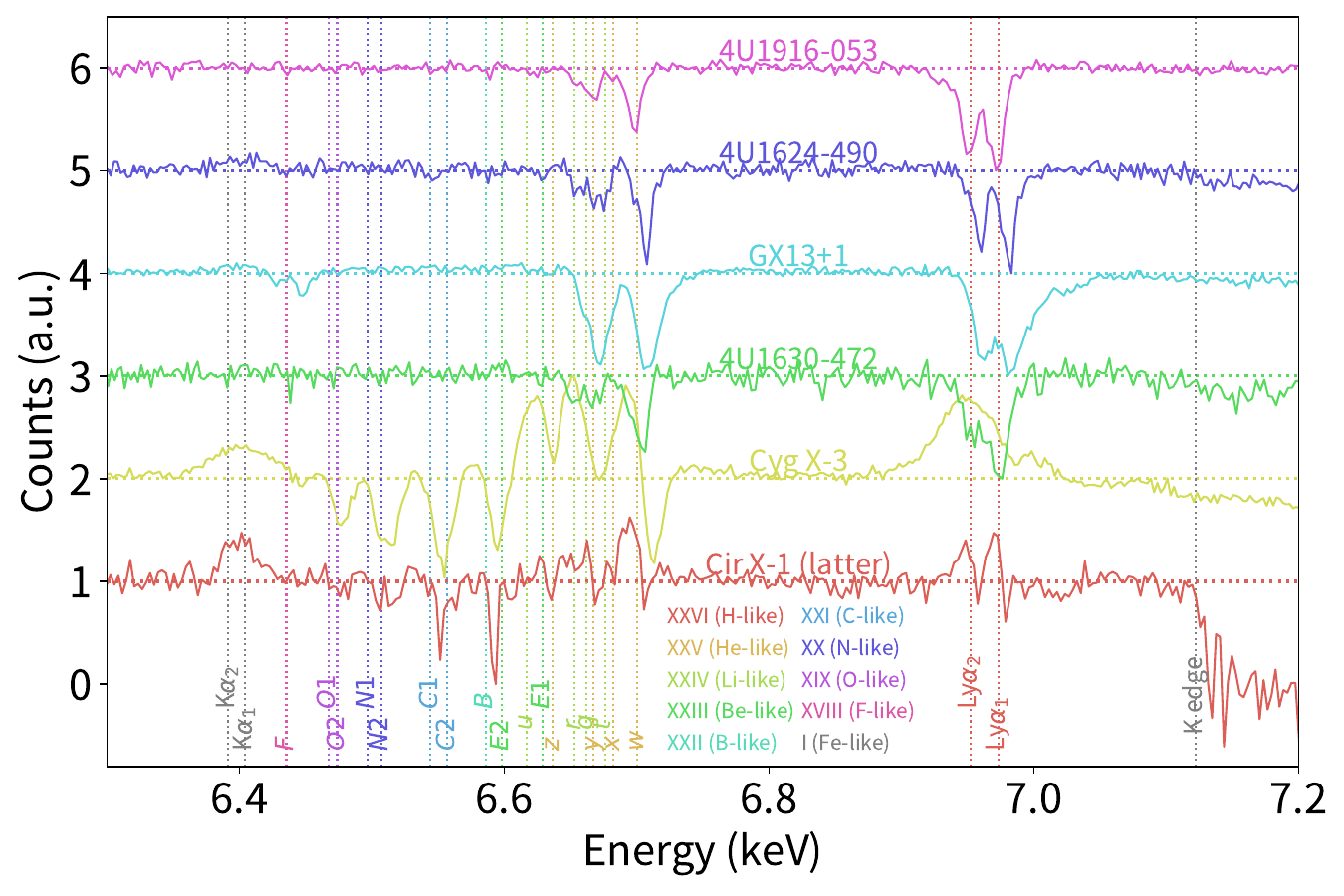}
 \end{center} 
 \caption{Fe K-band spectra of some X-ray binaries observed with
 \textit{Resolve}. Spectra are optimally-binned, detrended for the power-law continuum
 emission, scaled to $\pm$1, and offset for clarity. We use the average spectra; all
 sources exhibit spectral variability, the details of which are described in individual
 papers \citep{collaboration2024,xrismcollaboration2025,miller2025}. The rest-frame
 energies of Fe features are shown in different colors for Fe \emissiontype{I}
 (neutral), and \emissiontype{XVIII} (F-like) to \emissiontype{XXIV} (H-like). The
 labels follow \citet{rudolph2013}. The energy shift seen in most spectral features are
 real. {Alt text: a line plot showing X-ray spectra of some X-ray binaries observed with
 \textit{Resolve}.}}
 \label{f10}
\end{figure*}

Among the three parameters, $\xi$ is the most important governing the charge and level
populations and the formation of spectral features. We examined the calculation results
for different $\xi$ to compare to the observed spectra, focusing on the Fe emission
and absorption features. Figure~\ref{f12} shows the (a) charge fraction, (b) intensity
of the emission lines, and (c) optical depth of the absorption lines as a function of
$\xi$.

For the outer-shell transitions (Fe \emissiontype{XXVI} and \emissiontype{XXV}), the
emission and absorption lines of the same spectral feature probe plasma in different
$\xi$ ranges. This is because the emission lines are mostly formed via recombination
cascades, thus their intensity reflects the charge population that captures an electron;
e.g., Fe \emissiontype{XXVI} Ly$\alpha$ emission intensity reflects the fully-ionized
Fe$^{26+}$ ion population. In contrast, the absorption lines reflect the charge
population that absorbs a photon of the energy; e.g., Fe \emissiontype{XXVI} Ly$\alpha$
absorption depth reflects the H-like Fe$^{25+}$ ion population. Forbidden lines such as
Fe He$\alpha$ ($z$) are seen only in emission and not in absorption because the upper
level is mostly populated by the recombination cascade or inner-shell photoionization.

For the inner-shell transitions (Fe \emissiontype{XVIII} to \emissiontype{XXIV}), the
interpretation requires help from atomic physics theory and experiment. Their upper
levels are mostly populated by by photo-excitation or photo-ionization rather than by
electron collision or by dielectronic recombination, as the electron temperature is too
low to cause the latter two processes. Therefore, many absorption features are formed at
$\log{\xi}=$2--3, at which mildly-ionized Fe ions are dominant (figure~\ref{f12}c). For
elements of low atomic numbers, such photo-excited states decay predominantly by
auto-ionization ejecting electrons (Auger process) rather than radiative deexcitation
emitting photonhi (fluorescence). Therefore, they do not contribute for emission
lines. For elements of high atomic numbers like Fe, however, the fluorescent decay
becomes more dominant than the Auger decay, increasingly so for ions with a smaller
number of electrons. \citet{palmeri2003b} conducted \textit{ab initio} calculations of
the Dirac equation for such Fe ions and derived the radiative branching ratio from
various excited levels (table~\ref{t02}). The radiative branching ratio for the upper
level of the Fe \emissiontype{XXIV} $q$ line, for example, amounts to 1.00. Therefore,
they contribute significantly for the emission lines, and can form a P Cygni profile
similarly to outer-shell transitions. Such emission lines are indeed observed in a
ground experiment \citep{rudolph2013}. Mildly ionized Fe was collisionally ionized and
trapped by an electron beam ion trap (EBIT) installed at a synchrotron facility, which
provided a tunable, monochromatic photon beam that was passed through the ion
cloud. Prominent emission lines such as Fe \emissiontype{XXIV} $q$, $r$, Fe
\emissiontype{XXIII} $E1$ were observed as fluorescence decay of photo-excited states.

In photo-ionized plasmas around compact objects, these spectral features are predicted
by the atomic physics and radiative transfer combined, demonstrated by ground
experiments, and are now observed in the X-ray microcalorimeter spectra of Cir X-1
having a wide range of $\log{\xi}=$2--4 (figure~\ref{f06}). The uniqueness of Cir X-1 is
better understood by inspecting other X-ray binaries with the Fe K features. We compared
the spectra of Cyg X-3, 4U\,1630--472, GX\,13$+1$, 4U\,1624--490, and 4U\,1916--053
(sequence numbers 300065010, 900001010, 300036010, 300040010, and 300039010), which were
also observed with \textit{Resolve} as performance verification targets
(figure~\ref{f10}). Cir X-1 and Cyg X-3 show (i) absorption features by both mildly and
highly ionized Fe and (ii) emission features of highly ionized Fe, which are not the
case for the other sources. Furthermore, Cir X-1 exhibits the Fe Ly$\alpha$ absorption
at 7.0~keV as strong as the He$\alpha$ absorption at 6.7~keV. These characteristics are
the consequences of having plasma in wide $\log{\xi}$ ranges. The presence of the deep
absorption edge by Fe K$\alpha$ at 7.12~keV is also unique for Cir X-1.

\subsection{Line ratio analysis}\label{s5-2}
Before fitting the entire spectra, we estimate the plasma parameters using line
ratios. This is a necessary step for identifying approximate values of the parameters;
otherwise, the fitting often ends with a local minimum and ignores informative line
ratios because the $\chi^2$ statistics are dominated by spectral bins only with continuum
emission that are more numerous. For the $\xi$ values, we can constrain using the ratio
between the He$\alpha$ and Ly$\alpha$ resonance lines of each element
(\S\ref{s5-2-2}). However, we should be aware that these lines, in particular Fe, are
almost certainly optically thick as calculated in \S\ref{s1}. Therefore, we first
determine $N_{\mathrm{H}}$ (\S\ref{s5-2-1}).

\begin{figure}[!hbtp]
 \begin{center}
 \includegraphics[width=0.9\columnwidth,clip]{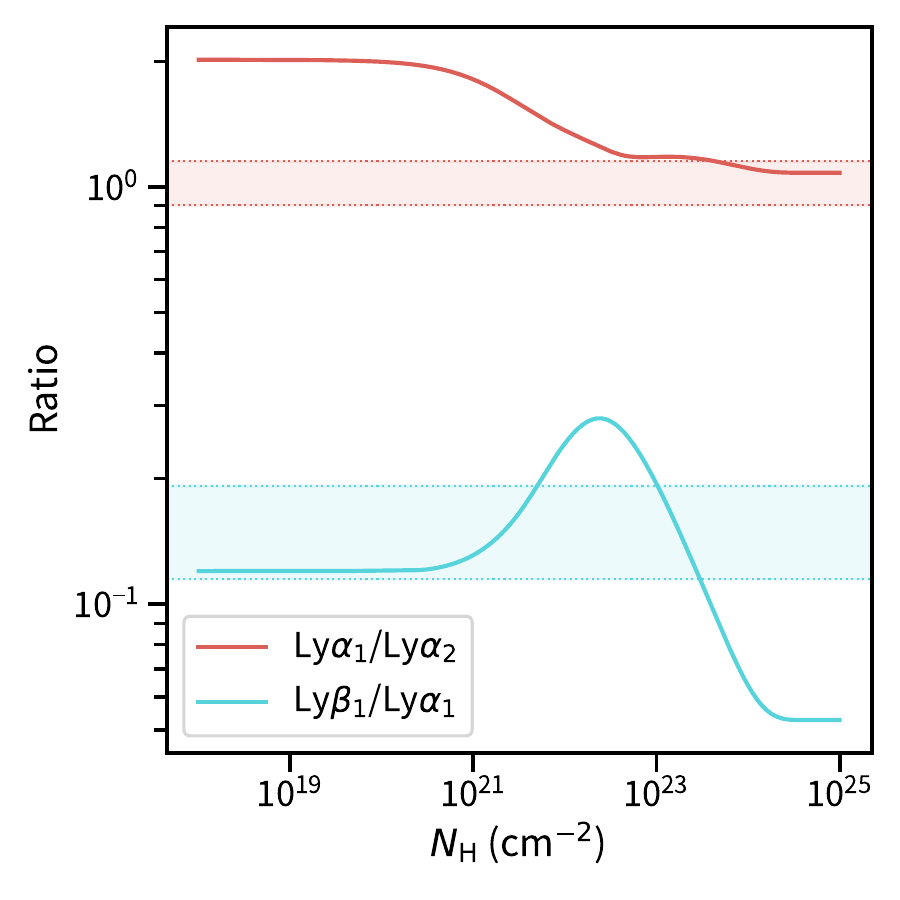}
\end{center}
 \caption{Line ratio of the Fe Ly$\alpha$ doublet (Ly$\alpha_1$/Ly$\alpha_2$) and Lyman
 decrement (Ly$\beta_1$/Ly$\alpha_1$) as a function of $N_{\mathrm{H}}$ calculated for
 $\log{\xi}=4.0$ (curves) and compared to the observed ratio (stripes) in the same
 colors. The M1 transition 2s~$^{2}S_{1/2}$$\rightarrow$1s~$^{2}S_{1/2}$ was
 blended into Ly$\alpha_2$. {Alt text: a line plot showing the line intensity ratio as a
 function of the plasma thickness.}}
 \label{f11}
\end{figure}

\begin{figure}[!hbtp]
  \begin{center}
   \includegraphics[width=1.0\columnwidth,clip]{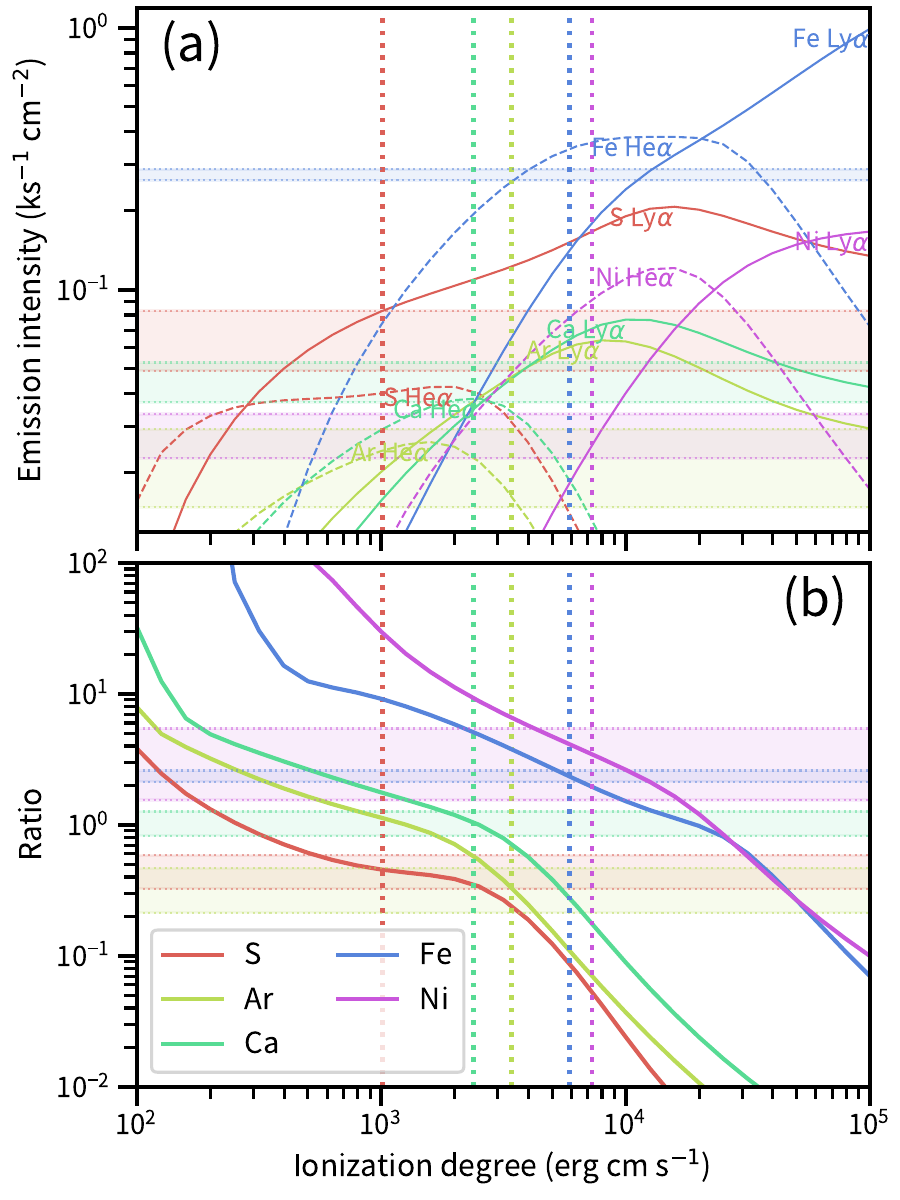}
  \end{center}
  \caption{(a) Emission line intensity as a function of $\xi$ for the resonance lines of
  H-like (solid) and He-like (dashed) S, Ar, Ca, Fe, and Ni lines for
  $N_{\mathrm{H}}=10^{23}$~cm. The emission line is scaled in the same way as
  figure~\ref{f12}. The observed 1$\sigma$ ranges of the He$\alpha$ ($w$) intensity are
  shown with the horizontal shades.  (b) Line ratio of He$\alpha$ ($w$) and Ly$\alpha_1$
  as a function of $\xi$. The observed 1$\sigma$ ranges of the line ratio are shown with
  the horizontal shades and their best matched $\xi$ value is shown in (a) and (b) with
  the vertical dotted lines of the same color for each element. {Alt text: two line
  plots showing emission line intensity and the intensity ratio of highly-ionized
  ions.}}
  \label{f13}
 \end{figure}

\begin{figure}[!hbtp]
 \begin{center}
  \includegraphics[width=0.9\columnwidth,clip]{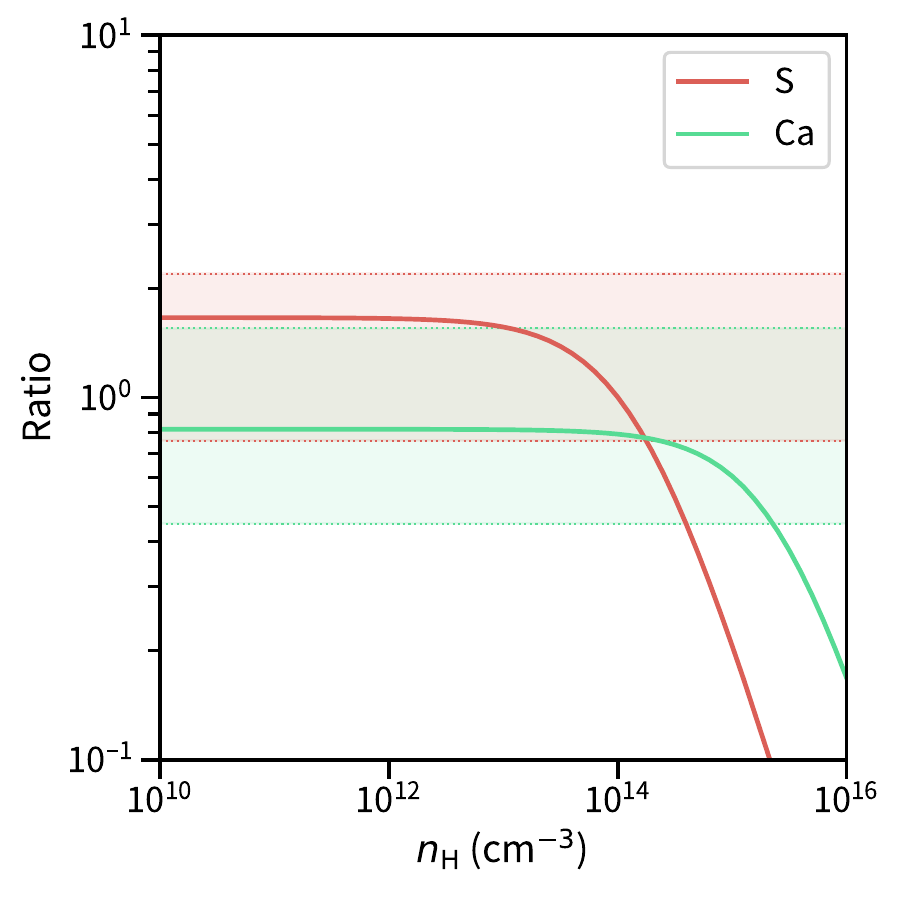}
 \end{center}
 \caption{$R$ ratio of S and Ca He$\alpha$ complex calculated for $\log{\xi}=3.2$ and
 $\log{N_{\mathrm{H}}}=22$. The observed ratios, using the entire dataset, are shown
 with stripes in the same color. {Alt text: a line plot showing the R ratio as a
 function of plasma density.}}
 \label{f16}
\end{figure}

\subsubsection{Column density}\label{s5-2-1}
A novel technique was recently proposed \citep{gunasekera2025} using the Fe Ly$\alpha$
fine-structure doublet, which requires the spectral resolution of \textit{Resolve}. The
ratio of the two emission lines (Ly$\alpha_1$/Ly$\alpha_2$) is 2 at the optically-thin
limit for their statistical weights, which decreases toward 1 as the line optical depth
increases, approaching the ratio of the source function at the depth where the line
optical depth reaches $\sim$2/3 seen from outside for each line
\citep{barbier1943}. Figure~\ref{f11} shows the result. Compared to the observation
(table~\ref{t01}), we estimate $\log{N_{\mathrm{H}}} \gtrsim 23$.

This can be further constrained using another pair of lines from the same ions
\citep{chakraborty2020c} such as Ly$\beta_1$ over Ly$\alpha_1$, which is called the
Lyman decrement in literature. The intensity ratio against $N_{\mathrm{H}}$ exhibits a
non-monotonic curve. Similarly to the Ly$\alpha$ doublet ratio, the Lyman decrement
approaches 1 as the line optical depth increases. However, it goes to the case D
condition, in which a Lyman $\beta$ photon degrades into a Balmer $\alpha$ photon and a
Lyman $\alpha$ photon during their repeated resonance scattering. Therefore, the line
ratio turns to a decreasing trend toward 0. By combining these diagnostics, we estimated
$N_{\mathrm{H}} \sim 23.5$.

\subsubsection{Ionization degree}\label{s5-2-2}
We next constrain $\xi$ using the He$\alpha$ and Ly$\alpha$ line ratio for the given
$N_{\mathrm{H}}$. We used S, Ar, Ca, Fe, and Ni in the former half of the
observation. Figure~\ref{f13} (a) shows the expected line intensity based on the RT
calculation. Figure~\ref{f13} (b) shows the intensity ratio of the pair compared to the
observation (table~\ref{t01}). We see that different elements trace different ranges of
$\log{\xi}=2.5-4.5$.

\subsubsection{Plasma density}\label{s5-2-3}
In collisionally-ionized plasmas \citep{gabriel69}, the $R \equiv I_{z}/(I_{x}+I_{y})$
ratio is often used as a density diagnostic, in which $I_{x}$, $I_{y}$, and $I_{z}$ are
the intensity of the $x$, $y$, and $z$ lines respectively in the He$\alpha$ line
complex. The upper levels of $x$ and $y$ lines are populated due to excitation by proton
and electron collisions and UV photon absorption from the upper level of $z$ before it
radiatively decays via a forbidden transition.

The $R$ ratio serves as a density indicator even in photoionized plasmas where radiative
processes dominate over collisional processes, but it is strongly affected by $\xi$ and
$N_{\mathrm{H}}$ \citep{porter2007}. In practice, we should also be aware that the
complex is often contaminated by other lines. To take Ar He$\alpha$ as an example, the
$z$ line blends with the S Ly$\beta$ line only at 3~eV away, which is serious
considering the presence of the S Ly$\gamma$ line detected at a comparable intensity
(figure~\ref{f06}b). We thus need to have the diagnostics based on dedicated RT
calculation. We used less contaminated S and Ca He$\alpha$ complexes and used the sum of
the former and the latter half of the spectrum for the sake of statistics. We obtained
an upper limit $\log{n} \sim 14$ from S (figure~\ref{f16}) consistent with previous
studies using the same technique \citep{Schulz2002,Iaria2008}.

\subsection{Spectral fitting}\label{s5-3}
\begin{figure}[!hbtp]
 \begin{center}
  \includegraphics[width=0.95\columnwidth,clip]{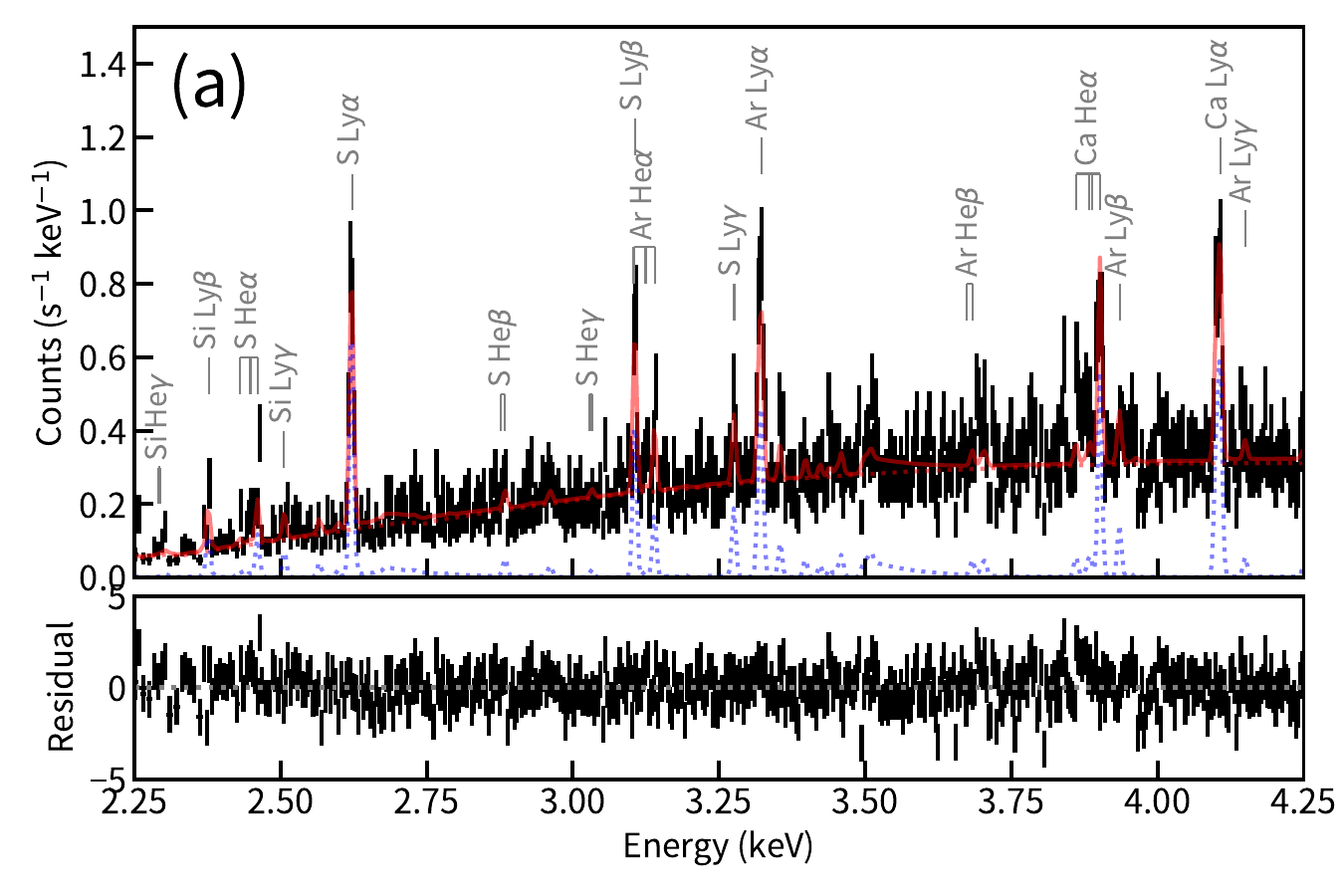}
  \includegraphics[width=0.95\columnwidth,clip]{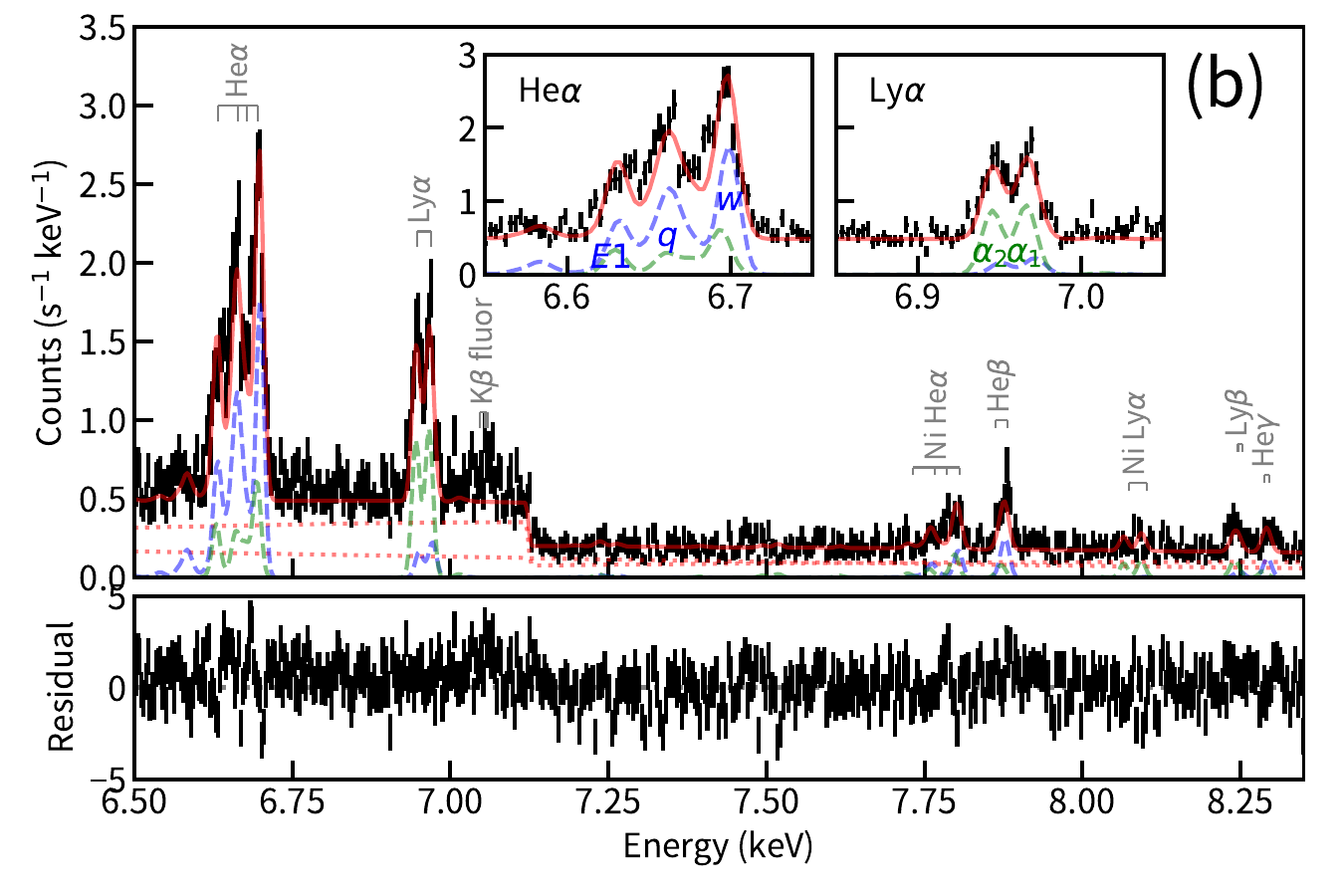}
  \includegraphics[width=0.95\columnwidth,clip]{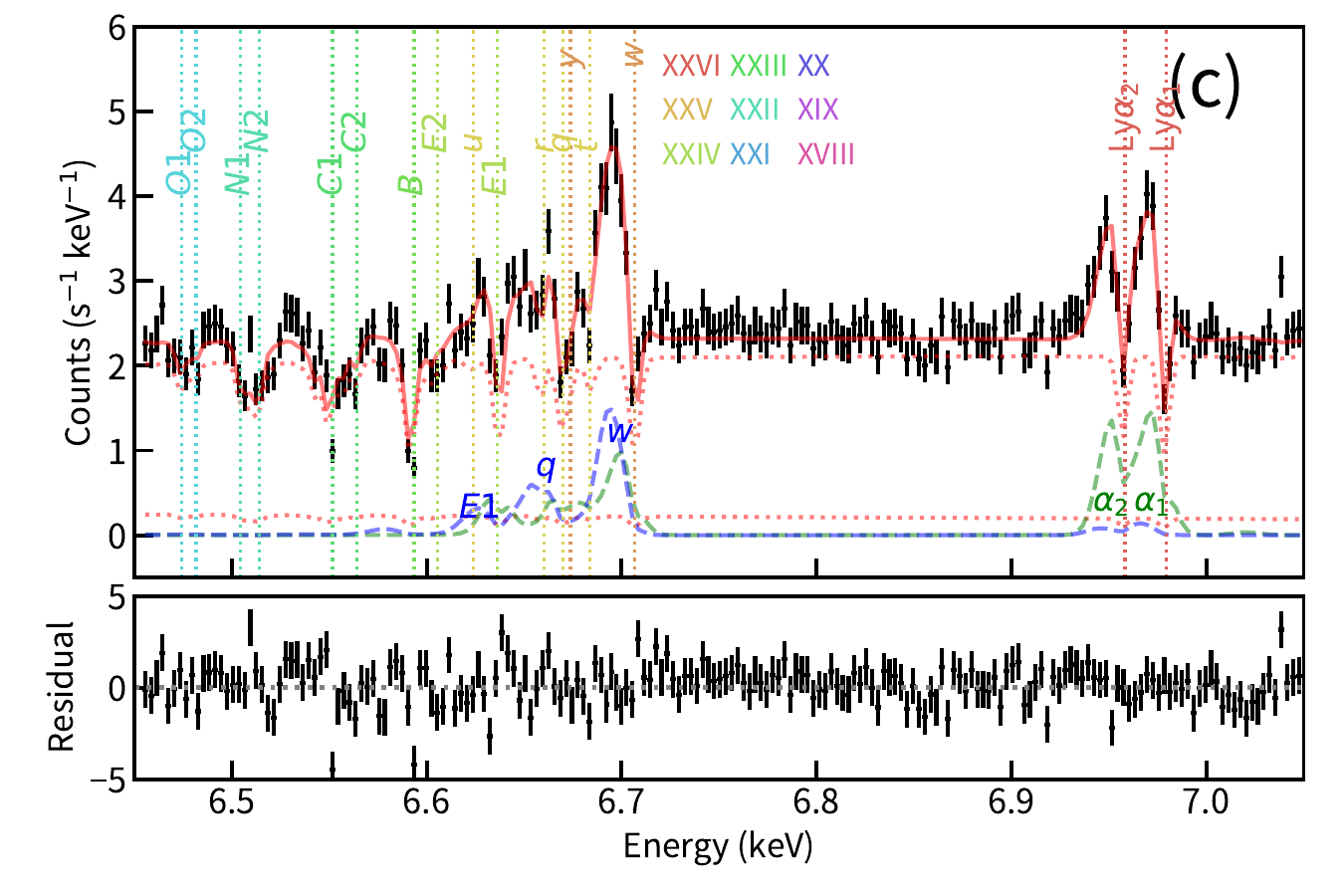}
 \end{center}
 \caption{Best-fit spectral model and data (upper panels) and the residuals (lower
 panels) for (a) former half, soft band, (b) former half, hard band, and (c) latter
 half, Fe K$\alpha$ band of the spectrum. Spectral model components are shown in solid
 (total), broken (transmitted and scattered), or dashed (diffuse) curves. Blue and green
 curves are respectively for the ``emi1'' and ``emi2'' component in table~\ref{t04}. The
 line labels are shifted by the best-fit Doppler velocity.  {Alt text: three line plots
 showing X-ray spectra and best-fit models.}}
 \label{f08}
\end{figure}

We now have a good grasp of the plasma parameters through the line ratio analyses. We
construct a spectral model to explain the observed spectra. We start with the former
half of the spectrum to constrain the plasma by its emission. Upon the best-fit model
for the continuum plus Fe K$\alpha$ fluorescence emission for the former half
(\S\ref{s4-2-1}), we added diffuse emission components of different $\xi$,
$N_{\mathrm{H}}$, emission measure (EM), and redshift ($v$) for the bulk motion. We note
that $N_{\mathrm{H}}$ and EM can be determined independently as we can constrain
$N_{\mathrm{H}}$ from the line ratios (figure~\ref{f11}). In the soft band
(2.25--4.25~keV; figure~\ref{f08}a), a single component with $\log{\xi} = 3.2$ suffices
to explain the S, Ar, and Ca emission lines.

In the hard band (6.45--8.5~keV; figure~\ref{f08}b), another component with a larger
$\log{\xi}$ value is required to explain the Fe and Ni emission lines. To explain the Fe
He$\alpha$ and Fe Ly$\alpha$ complex (insets in figure~\ref{f08}b), there are two
solutions. One is to explain both features with a single $\log{\xi} \sim 3.7$ component
as indicated in figure~\ref{f13} (b). In this case, the two peaks in the Ly$\alpha$
complex and the three peaks in the He$\alpha$ complex are all explained by this
component. The $w$, $y$, and $z$ lines are for the three peaks in the He$\alpha$
complex, as initially guessed in table~\ref{t01}. The other is to explain the Ly$\alpha$
complex with the larger $\xi$ component and the He$\alpha$ complex with the smaller
$\xi$ component. The $w$, $q$, and $E1$ lines are for the three peaks in the He$\alpha$
complex. We favor the latter, as the former requires different bulk velocities between
the Ly$\alpha$ and He$\alpha$ lines of the same component. The best-fit parameters are
given in table~\ref{t04}.

Finally, we used the latter half of the spectrum to constrain the plasma by its
absorption. We focus on the 6.45--7.05~keV band, at which most absorption features are
included. Upon the best-fit model for the continuum plus Fe K$\alpha$ fluorescence
emission for the latter half (\S\ref{s4-2-1}) and the emission line model for the former
half (figure~\ref{f08}b), we multiplied the plasma absorption models. As is evident from
the Fe Ly$\alpha$ line between the former and the latter half of the spectrum
(figure~\ref{f06}c), the absorption by the plasma needs to be multiplied to the emission
from the plasma. Two components with different $\xi$ values explain the observed
absorption features of Fe \emissiontype{XIX}--\emissiontype{XXVI} (figure~\ref{f08}c). The best-fit values are shown
in table~\ref{t04}.

\begin{table} 
 \tbl{Physical model fitting result.}{
 \begin{tabular}{lcccc}
  \hline
  Comp\footnotemark[$*$] & $\log{\xi}$ & $\log{N_{\mathrm{H}}}$ & EM & $v$\footnotemark[$\dagger$] \\
        & (erg~cm~s$^{-1}$) & (cm$^{-2}$) & (10$^{59}$ cm$^{-3}$) & (km~s$^{-1}$) \\
  \hline
  emi1 & $3.13^{+0.03}_{-0.03}$ & $22.4^{+0.0}_{-0.1}$ & $2.0 \pm 0.2$ & \phantom{0}$78^{+25}_{-10}$\\
  emi2 & $4.60^{+0.06}_{-0.09}$ & $24.2^{+0.2}_{-0.2}$ & $1.3 \pm 0.3$ & $337^{+19}_{-18}$\\
  \hline
  abs1 & $2.60^{+0.02}_{-0.01}$ & $23.0^{+0.0}_{-0.0}$ & --- & $-330^{+11}_{-7}$\\
  abs2 & $4.34^{+0.23}_{-0.20}$ & $22.4^{+0.2}_{-0.2}$ & --- & $-272^{+18}_{-16}$\\
  \hline
 \end{tabular}}
 \label{t04} 
 \begin{tabnote} 
 \footnotemark[$*$] ``emi1'' and ``emi2'' are the components to explain the emission
  lines in the former half (figure~\ref{f08}a and \ref{f08}b). ``abs1'' and ``abs2'' are the
  components to explain the absorption lines in the latter half (figure~\ref{f08}c). \\
 \footnotemark[$\dagger$] Energy shift from the fitting. The redward shift is positive.\\
 \end{tabnote} 
\end{table}

\section{Discussion}\label{s6}
We were successful in explaining the observed spectra with a physically motivated model
based on the RT calculation. We now interpret this based on the picture in
figure~\ref{f00}. We first discuss the cause of the observed X-ray variability in
\S~\ref{s6-1}. We then discuss the structure (\S~\ref{s6-2}) and dynamics
(\S~\ref{s6-3}) of the outflow based on the parameters constrained by comparing the
observation and RT calculations in \S~\ref{s5}.

\subsection{Covering materials}\label{s6-1}
The variability of the former and the latter half of the spectrum is mostly attributable
to the change of the normalization of the disk blackbody emission
(table~\ref{t03}). There are two possibilities. One is that the intrinsic normalization
of the emission has changed accordingly. The other is that the intrinsic normalization
did not change much, but the fraction of coverage by a very thick material in the line
of sight has changed. The material is so thick ($\gtrsim 10^{26}$~cm$^{-2}$) that almost
no emission comes out in the \textit{Resolve} energy band. We consider the latter to be
the case for three reasons. One is that the scattered emission, which we consider to be
the origin of the continuum emission below $\sim$4~keV, did not change. If the intrinsic
normalization of the incident emission had changed, the scattered emission would have
changed proportionally. The second is that the best-fit normalization of the disk
blackbody emission is too small. For the face value of $N^{\mathrm{(disk)}}=16$
(table~\ref{t03}) at a 9.4~kpc distance and $\theta=75$~degree, the inner disk radius is
$R_{\mathrm{in}}=7.7$~km, which is smaller than the typical NS radius
\citep{miller2019}. The third is that we see both emission and absorption features in
the latter half of the spectrum. If we had observed all the incident emission, the
transmitted emission through the plasma should overwhelm the diffuse emission from the
plasma, and we would have observed a spectrum predominantly characterized by absorption
features. This is the situation for the four sources from the top in figure~\ref{f10}.

In the former half, the thick material blocked a significant fraction of the incident
emission, and the diffuse emission dominates. This is akin to the situation of X-ray
binaries during eclipses or deep dips. Indeed, high-resolution X-ray spectra of such
sources are dominated by emission lines with little absorption features
\citep{schulz2002a,hirsch2019,pradhan2024a}, as we saw in figure~\ref{f06} (a).

When the thick material goes out of the line of sight, the dip phase ends
(figure~\ref{f02} inset), and we observe the entire intrinsic normalization of the
incident emission, which is assessed as $N^{\mathrm{(disk)}} \sim$ 40 at the peak with
the NICER spectrum \citep{tominaga2023}. This corresponds to $7.6 \times
10^{37}$~erg~s$^{-1}$ in 1--1000~Ryd. The scattered emission (table~\ref{t03}) has
$\sim$1\% of the intrinsic incident luminosity, suggesting that the optical depth of
electron scattering is $\sim 10^{-2}$, so $N_{\mathrm{H}} \sim
10^{-2}/\sigma_{\mathrm{es}}= 10^{23}$~cm$^{-2}$. This is in reasonable agreement with
the estimate from the line ratio analysis (\S~\ref{s5-2-1}).

The putative thick ($\gtrsim 10^{26}$~cm$^{-2}$) material and the observed PCA are in
the line of sight only during the dip phase along the orbital phase. The most likely
entity for them is the hot spot formed at the position where the accreting matter lands
on the accretion disk (figure~\ref{f00}). It may have some bimodal density structure
along its radius responsible for both the thick material and the PCA.

\subsection{Structure of outflow}\label{s6-2}
We constrain the system parameters for the geometry in figure~\ref{f00}. The inner
radius is estimated as $r_{\mathrm{in}} = \sqrt{L_{\mathrm{X}}/n\xi} = 9 \times 10^{10}
n_{12}$~cm, in which $n_{12}$ is the density in the unit of $10^{12}$~cm$^{-3}$. Then,
$r_{\mathrm{out}} = r_{\mathrm{in}} + N_{\mathrm{H}}/n =
(0.9n_{12}^{-\frac{1}{2}}+n_{12}^{-1}) \times 10^{11}$~cm. This is in reasonable
agreement with the scaling used in figures~\ref{f12} and \ref{f13} to match the RT
calculation with the observed line flux. This is a minimum estimate of
$r_{\mathrm{out}}$ by the assumption that the volume filling factor is unity for the
spherically symmetric assumption. We do not know the outer radius of the accretion disk,
but it should be smaller than the effective Roche-lobe radius of the binary of
10$^{12}$~cm \citep{eggleton83}. From these estimates, the launching radius of the
outflowing plasma is close to the outer edge of the accretion disk.  No substantial
photoionized plasma is present much closer to the NS than this, as we would otherwise
have observed the Fe Ly$\alpha$ emission line stronger than He$\alpha$ for a larger
$\xi$ by recombination of Fe$^{26+}$ ions (figures~\ref{f06}c and \ref{f13}). Likewise,
no substantial plasma is present much farther than this, as we would otherwise have
observed deeper absorption lines of Fe \emissiontype{XIX} ($O1$, $O2$) than those of Fe
\emissiontype{XX} ($N1$, $N2$) and Fe \emissiontype{XXI} ($C1$, $C2$) for a smaller
$\xi$ (figures~\ref{f12}c and \ref{f08}c).


\subsection{Dynamics of outflow}\label{s6-3}
The Eddington luminosity for a $1.4~M_{\odot}$ NS is $L_{\mathrm{Edd}} = 1.8 \times
10^{38}$~erg~s$^{-1}$. At the time of the present observation, Cir X-1 was at a
sub-Eddington luminosity $L_{\mathrm{X}}= 7.6 \times 10^{37}$~erg~s$^{-1}$ $\sim 0.4
L_{\mathrm{Edd}}$ (\S~\ref{s4-2-1}), but an outflowing wind was observed with a
peak-to-valley distance in the P Cygni profile of $\sim 5\times 10^{2}$~km~s$^{-1}$
(\S~\ref{s4-2-2} and \S~\ref{s4-2-3}). One would naturally expect that the line photons
contribute for the outward radiative force in addition to the continuum photons. The
enhancement is described by the radiative force multipiler $\mathcal{M} \equiv
(\sigma_{\mathrm{es}}+\langle\sigma_{\mathrm{line}}\rangle)/\sigma_{\mathrm{es}}$, in
which $\langle\sigma_{\mathrm{line}}\rangle$ is the cross section of line photon
interactions averaged over wavelengths weighted by the incident spectral shape
\citep{tarter1973}. The Eddington luminosity is effectively reduced by
$L_{\mathrm{Edd}}/\mathcal{M}$. $\mathcal{M}$ decreases as $\xi$ increases because more
electrons are stripped from the ions. This is in line with the observation, in which the
blue-shifted velocity of 330$^{+7}_{-11}$~km~s$^{-1}$ for the smaller $\xi$ component is
significantly faster than that of 272$^{+16}_{-18}$~km~s$^{-1}$ for the larger $\xi$
component (table~\ref{t04}).

\begin{figure}[!hbtp]
 \begin{center}
  \includegraphics[width=1.0\columnwidth,clip]{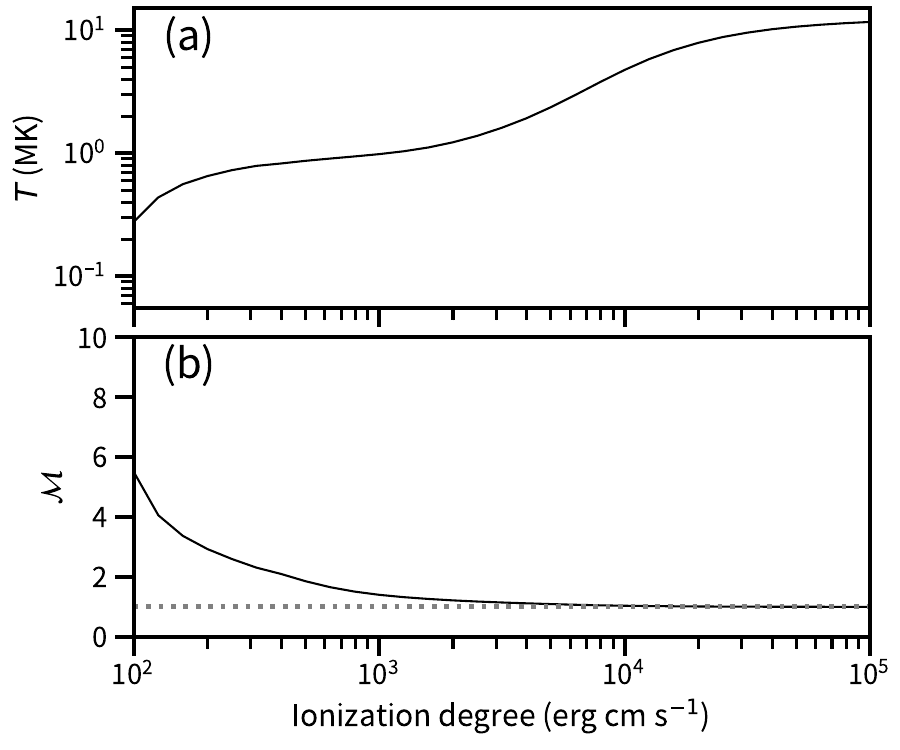}
 \end{center}
 \caption{(a) Plasma temperature ($T$) and (b) force multiplier $\mathcal{M}$ as a
 function of $\xi$ based on the RT calculation.}
 \label{f14}
\end{figure}

We can derive $\mathcal{M}(\xi)$ based on the presented RT calculation using
\texttt{cloudy}, as we have solved all the photon-matter interaction and verified it
with the observed X-ray spectra. Figure~\ref{f14} shows $\mathcal{M}(\xi)$ as
well as the plasma temperature $T(\xi)$ at which the heating and cooling balance. The
observed bulk velocity is larger than the sound velocity for the plasma temperature, and
the Sobolev condition holds for the line force to work efficiently. The general trend of
$\mathcal{M}(\xi)$ is the same as the classical work by \citet{tarter1973} among
variations of incident spectral shapes. For our setup of the disk blackbody emission
representative of LMXBs, $\mathcal{M(\xi)}>3$ at $\log{\xi} < 2.1$, which makes the
observed $L_{\mathrm{X}}$ to exceed $L_{\mathrm{Edd}}/\mathcal{M}(\xi)$ and can give
rise to radiatively-driven outflow. The faster component with a smaller $\xi$ may drag
the slower component with a larger $\xi$ if they spatially coexist. For the disk wind,
we would also expect that the centrifugal force outward also works to offset the
gravitational force inward. Therefore, we argue that the radiative force by continuum
and line photons is sufficient to produce the observed outflowing plasma. The total mass
loss rate, if radially symmetric, is estimated as $4 \pi r_{\mathrm{out}}^2 n
m_{\mathrm{p}} v \approx 2 \times 10^{-7}
(1+0.9\sqrt{n_{12}})M_{\odot}$~yr$^{-1}$.

\section{Conclusion}\label{s7}
The new era of high-resolution X-ray spectroscopy with X-ray microcalorimeters has
begun. Most spectral features in plasmas around compact objects are formed in NLTE
conditions, and thus RT calculation is required to interpret the data. We presented such
a study for a LMXB using \texttt{cloudy} focusing on line ratio analysis.

Many new spectral features, both emission and absorption, were detected with
\textit{Resolve} from Cir X-1 at $\phi_{\mathrm{orb}} =0.926-0.973$. The X-ray spectra
changed drastically in the former and the latter half of the observation, which was
explained by the apparent change in the continuum emission. The diffuse and scattered
emission from the photoionized plasma remained unchanged, suggesting that the X-ray
variability across the orbital phase is caused by the local intervening matter on the
accretion disk as proposed by \citet{tominaga2023}.

We constrained the plasma parameters ($\log{\xi} = 2.5-4.5$, $\log{N_{\mathrm{H}}} \sim
23$, and $\log{n} \lesssim 14$) using the line ratio analysis. These values are consistent with
the independent estimates using the electron scattering optical depth and absolute line
flux. We further derived the outward velocity ($\sim$300~km~s$^{-1}$) of the plasma and
its distance from the NS as $\mathcal{O}$($10^{11}$~cm), based on which we argue that
the launching radius of the outflowing plasma is close to the outer edge of the
accretion disk. We further calculated the radiative force multiplier based on the RT
calculation and argue that the radiative force, including the line force contribution,
is sufficient to drive the observed outflow at a sub-Eddington luminosity.

\begin{ack}
 The authors appreciate Gary J. Ferland at the University of Kentucky, Chamani
 M. Gunasekera at STSci, Peter van Hoof at the Royal Observatory of Belgium, and
 Stephano Bianchi at Universit\`{a} degli Studi Roma Tre for their efforts to make
 \texttt{cloudy} useful for X-ray spectroscopy as presented here and their expertise
 advice since a \texttt{cloudy} summer school held at ISAS in 2024. We also appreciate
 Mayu Tominaga for her contribution in scientific justification of this performance
 verification program and observation planning through her thesis. We thank all those
 who contributed to the XRISM mission. This work was supported by the JSPS Core-to-Core
 Program (grant number: JPJSCCA20220002). MS acknowledges support by Grants-in-Aid for
 Scientific Research 19K14762 and 23K03459 from the Ministry of Education, Culture,
 Sports, Science, and Technology (MEXT) of Japan. Part of this work was performed under
 the auspices of the U.S. Department of Energy by Lawrence Livermore National Laboratory
 under Contract DE-AC52-07NA27344. The material is based upon work supported by NASA
 under award number 80GSFC21M0002. This research made use of the JAXA's high-performance
 computing system JSS3.
\end{ack}

\bibliography{ms}{}
\bibliographystyle{aasjournal}
\end{document}